\PassOptionsToPackage{unicode}{hyperref}
\PassOptionsToPackage{naturalnames}{hyperref}
\documentclass[preprint,3p,times]{elsarticle}

\usepackage{amssymb}
\usepackage{amsmath}

\usepackage{etex}
\usepackage{graphicx}
\usepackage{amsmath}
\usepackage{amssymb}
\usepackage{amsthm}
\usepackage{empheq}
\usepackage{enumitem}
\usepackage{multirow}
\usepackage{makecell}
\usepackage{siunitx}
\usepackage{pict2e}
\usepackage{booktabs}
\usepackage[figuresright]{rotating}
\usepackage{bbm}
\usepackage{mathtools}  
\usepackage{xfrac}
\usepackage{tikz}
\usepackage[abs]{overpic}
\usepackage{xcolor,varwidth}
\usetikzlibrary{angles,quotes}
\tikzset{>=latex} 
\usepackage{xcolor}
\colorlet{veccol}{green!70!black}
\colorlet{vcol}{green!70!black}
\colorlet{xcol}{blue!85!black}
\colorlet{projcol}{xcol!60}
\colorlet{unitcol}{xcol!60!black!85}
\colorlet{unitcol2}{vcol!60!black!85}
\colorlet{myblue}{blue!70!black}
\colorlet{myred}{red!70!black}
\tikzstyle{vector}=[->,very thick,xcol]
\tikzstyle{mydashed}=[dash pattern=on 2pt off 2pt]

\usepackage{nomencl}   

\usepackage{textcomp} 
\usepackage{pifont}   
\usepackage{booktabs}
\usepackage{physics}
\usepackage{wrapfig}
\usepackage{subcaption}
\usepackage{algorithm}      
\usepackage{pgfplots}
\pgfplotsset{compat=1.17}
\usepackage{listings}		
\usepackage[noend]{algpseudocode}
\usepackage{changepage}     
\usepackage{color}              
\definecolor{myred}{gray}{0} 
\definecolor{light-gray}{gray}{0.8}
\usepackage{epstopdf}
\usepackage{comment}
\usepackage{graphics} 
\usepackage{epsfig} 
\usepackage{pdflscape}
\usepackage{overpic}

\usepackage{lineno}

\newtheorem{theorem}{Theorem}[section]
\newtheorem{definition}{Definition}[section]

\newtheorem{proposition}{Proposition}[section]

\newtheorem*{theorem*}{Theorem}
\newtheorem*{corollary*}{Corollary}
\newtheorem*{lemma*}{Lemma}
\newtheorem*{hypothesis*}{Hypothesis}
\newtheorem*{proposition*}{Proposition}
\newtheorem*{conjecture*}{Conjecture}

\newtheorem{remark}[theorem]{Remark}

\theoremstyle{definition}
\theoremstyle{remark}
\theoremstyle{algorithm}

\let\originalleft\left
\let\originalright\right
\renewcommand{\left}{\mathopen{}\mathclose\bgroup\originalleft}
\renewcommand{\right}{\aftergroup\egroup\originalright}

\binoppenalty=\maxdimen 
\relpenalty=\maxdimen   
\numberwithin{equation}{section}    
\date{\today}             

\journal{Nuclear Physics B}

\begin{document}

\begin{frontmatter}



\title{Model-Adaptive Simulation of Hierarchical Shallow Water Moment Equations in One Dimension}


\author[rug]{Rik Verbiest} 
\author[rug,ug]{Julian Koellermeier}

\affiliation[rug]{organization={Bernoulli Institute, University of Groningen},
            addressline={Nijenborgh 9}, 
            city={Groningen},
            postcode={9747 AG}, 
            state={},
            country={The Netherlands}}

\affiliation[ug]{organization={Department of Mathematics, Computer Science and Statistics, Ghent University},
	addressline={Krijgslaan 281-S9}, 
	city={Ghent},
	postcode={9000}, 
	state={},
	country={Belgium}}

\begin{abstract}
Shallow free surface flows are often characterized by both subdomains that require high modeling complexity and subdomains that can be sufficiently accurately modeled with low modeling complexity. Moreover, these subdomains may change in time as the water flows through the domain. This motivates the need for space and time adaptivity in the simulation of shallow free surface flows. In this paper, we develop the first adaptive simulations using the recently developed Shallow Water Moment Equations, which are an extension of the standard Shallow Water Equations that allow for vertically changing velocity profiles by including additional variables and equations. The model-specific modeling complexity of a shallow water moment model is determined by its order. The higher the order of the model, the more variables and equations are included in the model. Shallow water moment models are ideally suited for adaptivity because they are hierarchical such that low-order models and high-order models share the same structure. To enable adaptive simulations, we propose two approaches for the coupling of the varying-order shallow water moment equations at their boundary interfaces. The first approach dynamically updates padded state variables but cannot be written in conservative form, while the second approach uses fixed padded state variable values of zero and reduces to conservative form for conservative moment equations. The switching procedure between high-order models and low-order models is based on a new set of model error estimators, originating from a decomposition of the high-order models. Numerical results of a dam break and of the collision of a dam-break wave with a smooth wave for different friction parameter values yield accurate results, while achieving speedups up to 40 percent compared to a non-adaptive model with fixed modeling complexity.
\end{abstract}


\begin{highlights}
\item First model-adaptive shallow flow simulation using hierarchical moment models.
\item New model-error estimators decompose the domain into dynamic varying-order subdomains. 
\item Two buffer cell-based spatial coupling formulas yield stable subdomain coupling.
\item Our new model-adaptive scheme achieves up to 40\% speedup in dam-break tests.
\end{highlights}


\end{frontmatter}



\section{Introduction}\label{section:Introduction}

The Shallow Water Equations (SWE) are nonlinear partial differential equations for flows with horizontal length scales much larger than the vertical one. They are derived by depth-integrating the incompressible Navier-Stokes equations under the shallowness assumption and the hydrostatic pressure approximation, using the depth-averaged horizontal velocity as the primary unknown. However, the classical SWE do not represent the vertical structure of the velocity field explicitly, and therefore filter out higher-order effects such as vertical shear and dispersion \cite{kern2009measured}. This severe simplification renders the classical SWE inaccurate in applications such as dam breaks, where the vertical variability cannot be neglected \cite{inaccuracy_dambreak}. For this reason, the Shallow Water Moment Equations (SWME) were derived in \cite{SWME}. These hierarchical equations are more flexible than the classical SWE, because they allow vertical variability in the horizontal velocity components by introducing \(M\in\mathbb{N}\) additional variables, so-called moments, that describe the vertical velocity profile of the horizontal velocity components at the additional complexity of adding also \(M\) additional evolution equations for those moments. The SWME have successfully been applied to 1D shallow free surface flows \cite{HSWME,SWME}, open curved shallow channel flow \cite{Steldermann_2023}, radially symmetric free surface flow \cite{ASWME_Verbiest} and bedload transport problems \cite{SWME_bedload}, for example.

The level of complexity included in the SWME model is determined by its order \(M\). On the one hand, a higher order \(M_{\mathrm{H}}\in\mathbb{N}\) means that more variables and more equations are included in the model, increasing the modeling complexity and increasing the accuracy of simulations with non-uniform flow \cite{HSWME,SWME,ASWME_Verbiest}, i.e., flow in which the horizontal velocity profile varies considerably over the vertical position variable. A lower order \(M_{\mathrm{L}}\in\mathbb{N}\), on the other hand, includes less variables and equations and is therefore computationally less expensive, at the cost of accuracy loss in non-uniform flow simulations. Thus, choosing the order \(M\) of the SWME model is related to a trade-off between accuracy and computational efficiency. More precisely, one wants to simulate a shallow fluid flow using the SWME with as few moments as possible while achieving acceptable accuracy. In simulations where the flow is complex, for example when the flow velocity is non-uniform, one can expect to need a high order to achieve acceptable accuracy for the user. On the other hand, a simulation of a near-uniform flow, i.e., a flow in which the horizontal velocity only has small variations in the vertical variable, can be described sufficiently accurately by a low-order SWME model, including the classical SWE \cite{garcia-navarro_shallow_2019}. Note that the SWME of order zero coincide with the classical SWE.

In many applications, the complexity of the flow varies throughout the flow domain. This is the case in dam breaks, for example, as the flow is discontinuous at the wave induced by the dam break, but the flow is smooth in that part of the flow that the dam break waves have not reached yet \cite{fondelli_numerical_2015,lakhlifi_dam-break_2018}. This calls for adaptivity in the flow simulation. Different methods exist to incorporate adaptivity. One common adaptive approach is applying the adaptive mesh refinement algorithm \cite{berger_adaptive_1984} to the SWE \cite{donat_well-balanced_2014}. For vertically resolved models, various adaptive strategies for the vertical discretization of fluid flows have been proposed in the literature, including discontinuous Galerkin schemes with local polynomial-degree adaptivity (so-called $p$-adaptivity) \cite{KUBATKO20091766}, adaptive vertical grids and node redistribution techniques for the vertical structure of ocean models \cite{HOFMEISTER201070} and shallow water models \cite{FORTUNATO1999639}, and multilayer shallow water models with locally variable number of layers \cite{BONAVENTURA2018209}.


A different approach to incorporate adaptivity in the simulation of the fluid is not to locally modify the numerical method but to locally change the model instead. We call this model adaptivity. The advantage of model adaptivity is that the local flow complexity can be directly incorporated into the local model. The key component is then a necessary coupling of the different models and the different numerical methods used to simulate the models, which may or may not be difficult, depending on the types of model and the types of numerical method used. The SWME, however, are ideally suited for this approach, because different-order SWME models share the same structure, such that the same numerical scheme can be used in the entire domain. Moreover, the SWME are hierarchical, as lower-order SWME are included as a special case in higher-order SWME, which simplifies the coupling of adaptive hierarchical SWME models considerably. Previous works have studied the coupling of the fully resolved Reynolds-Averaged Navier-Stokes equations with the SWE \cite{mintgen_coupling_2018} and with SWME models of fixed order \(M\) \cite{steldermann2025dimensionally}. So far, no adaptive method coupling two different-order SWME models exists. 

Adaptive simulation of moment models has recently also gained increased interest for the simulation of rarefied gases in kinetic theory using moment models. This inspires the development of adaptive simulation methods for the SWME as the kinetic moment models and the SWME have a similar form. In \cite{schmeiser_convergence_1998}, a heuristic procedure is used to vary the order of a kinetic moment model. Other authors base their domain decomposition on a direct estimation of the model error \cite{abdelmalik_error_2017}. In \cite{koellermeier_error_2019}, error estimators based on (gradients of) the numerical solution are numerically studied. 

Our proposed adaptive simulation strategy of the SWME is the following: in regions of high complexity, a high-order SWME model with order \(M_{\mathrm{H}}\) is used to capture the complexity with sufficient accuracy, while in regions of low complexity, a low-order SWME model with order \(M_{\mathrm{L}}\) with small computational effort is used with sufficient accuracy. This is similar to the concept of $p$-adaptivity in finite element methods, but in contrast to $p$-adaptive finite element methods, model-adaptive SWME locally modify the model hierarchy itself. The switching procedure between high-order SWME and low-order SWME should be based on a suitable set of error estimators that estimate the model error made using a SWME model of a certain order. In this paper, a set of error estimators based on a decomposition of the SWME is proposed.  


In this paper, the first model-adaptive procedure for the simulation of a free surface flow modeled by the SWME model is introduced. This will accomplish the goal to derive a generalizable numerical scheme for the accurate but fast adaptive simulation of the SWME. A path-conservative finite volume scheme is used to numerically solve the SWME model in non-conservative form. The adaptive procedure consists of the following two steps: (1) The first step decomposes the domain into subdomains, each modeled by SWME models with a time- and space-dependent order by locally evaluating a set of error estimators, constructed based on a decomposition of the SWME model; (2) The second step is the coupling of the different-order SWME models at their boundary interfaces in a stable way and (ideally) such that conserved quantities are conserved. The advantage of this approach is that each subdomain is modeled with sufficient physical complexity and that the same finite volume solver can be used in the entire domain. 

The remainder of the paper is organized as follows. In Section \ref{section:SWME}, the SWME model is briefly reviewed. The error estimators and the coupling of the subdomains are discussed in Section \ref{section:Introduction_SpatiallyAdaptiveSimulation}. In Section \ref{section:Numerics}, numerical simulations are performed to test the adaptive procedure. The paper ends with a brief conclusion. 

\section{Shallow Water Moment Equations}\label{section:SWME}

The SWME were derived in \cite{SWME} as an extension of the classical SWE. In the two-dimensional case (one horizontal dimension \(x\in\Omega_x\subset\mathbb{R}\) and one vertical dimension \(z\in\Omega_z\subset\mathbb{R}\)), the derivation starts from the conservation of mass and momentum given by the incompressible Navier-Stokes equations in the physical space \((x,z)\in\Omega_x\times \Omega_z\), describing the evolution of the velocity \((u,w)\in\mathbb{R}^2\), where \(u\) is the horizontal velocity and where \(w\) is the vertical velocity. Under the assumptions that the flow is shallow, i.e., the horizontal scale is much larger than the vertical scale, that the pressure is hydrostatic, that the fluid is Newtonian with kinematic viscosity \(\nu\in\mathbb{R}^+\), and that the bed is fixed and horizontal, the incompressible Navier-Stokes equations can be reduced to
\begin{align}\label{eq:NSE-1}
	&\partial_x u + \partial_z w = 0,\\ \label{eq:NSE-2}
	&\partial_t u + \partial_x \left( u^2 \right)  + \partial_z \left( uw \right) = -\frac{1}{\rho} \partial_x p +\nu\partial_{zz}^2 u, 
\end{align}
with hydrostatic pressure \(p(t,x,z)=\left(h_s(t,x)-z\right)\rho g\) satisfying a stress-free boundary condition \(p(t,x,h_s(t,x))=0\) at the free surface, where \(\rho\) and \(g\) are the density and the gravitational constant, respectively, and where \(h_s(t,x)\) is the free surface. In \cite{SWME}, the vertical variable \(z\) is transformed to 
\begin{equation}\label{eq:zeta}
	\zeta = \frac{z-h_b}{h(t,x)}, \qquad \zeta \in [0,1],
\end{equation}
where \(h(t,x)=h_s(t,x)-h_b\) is the water height, with constant bottom topography $h_b\in\mathbb{R}$. Equations \eqref{eq:NSE-1}-\eqref{eq:NSE-2} are transformed to
\begin{align}\label{eq:ref1}
	&\partial_t h + \partial_x\left( hu_m \right)=0, \\[5pt]\label{eq:ref2}
	&\partial_t\left( h\Tilde{u} \right) + \partial_x\left( h\Tilde{u}^2+\frac{g}{2}h^2 \right) + \partial_{\zeta}\left(h\Tilde{u}\omega-\frac{\nu}{h}\partial_{\zeta}\Tilde{u}\right) = 0.
\end{align}
The variables of interest are the water height \(h(t,x):[0,T]\times \Omega_x\mapsto \mathbb{R}^+\) and the transformed horizontal velocity \(\Tilde{u}(t,x,\zeta):[0,T]\times\Omega_x\times [0,1]\mapsto \mathbb{R}\). The mean velocity \(u_m(t,x):[0,T]\times \Omega_x\mapsto \mathbb{R}\) is defined by \(u_m(t,x):=\int_{0}^1 \Tilde{u}(t,x,\zeta)d\zeta\). Finally, \(\omega[h,\Tilde{u}]\) is the vertical coupling operator, defined by
\begin{equation}\label{eq:vertical-coupling}
	h\omega[h,\Tilde{u}] := -\partial_x\left(h\int_0^{\zeta}(u_m(t,x)-\widetilde{u}(t,x,\Hat{\zeta}))d\Hat{\zeta}\right).
\end{equation}
In the remainder of this paper, the tilde in \(\Tilde{u}\) is dropped for readability. The main idea in the derivation of the SWME \cite{SWME} is to expand the horizontal velocity \(u(t,x,\zeta)\) in a truncated Legendre sum of order \(M\):
\begin{equation}\label{eq:legendre-sum}
	u(t,x,\zeta) = u_m(t,x) + \sum_{i=1}^{M}\alpha_i(t,x)\phi_i(\zeta),
\end{equation}where \(\alpha_i(t,x):[0,T]\times \Omega_x\mapsto \mathbb{R}\) are the basis coefficients, also called moments, and the basis functions \(\phi_i(\zeta):[0,1]\mapsto \mathbb{R}\) are the scaled and shifted Legendre polynomials of degree \(i\), with \(i=1,\ldots,M\), where \(M\in\mathbb{N}\) is the order of the SWME model that determines how many basis coefficients and basis functions are included in the expansion. Equations for the moments \(\alpha_i(t,x)\) are obtained by projecting the momentum equation \eqref{eq:ref2} onto the Legendre polynomials \(\phi_i(\zeta)\), i.e., multiplication of \eqref{eq:ref2} with \(\phi_i(\zeta)\), \(i=1,\ldots,M\), and subsequent integration with respect to \(\zeta\). This is a generalization of depth-averaging, which is equivalent to projecting \eqref{eq:ref2} onto the constant test function \(\phi_0(\zeta)=1\). We refer the interested reader for details to \cite{SWME}. The resulting \(M\)th order SWME model reads
\begin{align}\label{eq:SWME-h}
	\partial_t h + \partial_x\underbrace{\left( hu_m \right)}_{=:F_0} &= 0, \\[5pt]\label{eq:SWME-hum}
	\partial_t\left( hu_m \right)+\partial_x\underbrace{\left(h\left(u_m^2+\sum_{i=1}^{M}\frac{\alpha_j^2}{2j+1}\right)+\frac{g}{2}h^2\right)}_{=:F_1} &= \underbrace{-\frac{\nu}{\lambda}\left( u_m+\sum_{i=1}^{M}\alpha_i \right)}_{=:S_1}, \\
		\label{eq:SWME-halphai}
		\partial_t\left(h\alpha_i\right) + \partial_x\underbrace{\left(h\left(2u_m\alpha_i+\sum_{j,k=1}^{M}A_{ijk}\alpha_j\alpha_k\right)\right)}_{=:F_{i+1}} &= \underbrace{u_m\partial_x\left(h\alpha_i\right) -\sum_{j,k=1}^{M}B_{ijk}\alpha_k\partial_x\left(h\alpha_j\right)}_{=:Q_{i+1}} \underbrace{-(2i+1)\frac{\nu}{\lambda}\left(u_m+\sum_{j=1}^{M}\left(1+\frac{\lambda}{h}C_{ij}\right)\alpha_j\right)}_{=:S_{i+1}},
\end{align}
\(i=1,\ldots,M\), where the constant coefficients \(A_{ijk}\), \(B_{ijk}\) and \(C_{ij}\) are given by
\begin{equation}
	A_{ijk} = (2i+1)\int_0^1 \phi_i(\zeta)\phi_j(\zeta)\phi_k(\zeta)d\zeta, \quad B_{ijk} = (2i+1)\int_0^1 \phi_i'(\zeta)\left(\int_0^\zeta\phi_j(\hat{\zeta})d\hat{\zeta}\right)\phi_k(\zeta)d\zeta, \quad C_{ij} = \int_0^1 \phi'_i(\zeta)\phi'_j(\zeta)d\zeta,
\end{equation}
and where \(\lambda \in \mathbb{R}^+\) is the slip length, originating from the slip boundary condition at the bottom. Collecting the conservative fluxes in the vector \(\vec{F}_M(\vec{w}_{M}):=(F_0,F_1,\ldots,F_{M+1})\in\mathbb{R}^{M+2}\), the non-conservative fluxes in the vector \(\vec{Q}_M(\vec{w}_M):=(0,0,Q_2,\ldots,Q_{M+1})\in\mathbb{R}^{M+2}\), the source terms in the vector \(\vec{S}_M(\vec{w}_M):=(0,S_1,\ldots,S_{M+1})\in\mathbb{R}^{M+2}\) and constructing the non-conservative flux matrix \(Q_M(\vec{w}_M)\in\mathbb{R}^{(M+2)\times(M+2)}\) from the non-conservative fluxes \(\vec{Q}_M(\vec{w}_M)\), Equations \eqref{eq:SWME-h}-\eqref{eq:SWME-halphai} can be written in compact form as
\begin{equation}\label{eq:SWME-compact}
	\partial_t \vec{w}_M + A_M(\vec{w}_M)\partial_x\vec{w}_M = \vec{S}_M(\vec{w}_M),
\end{equation}
with state variable vector 
\begin{equation}\label{eq:state-variable-vector}
	\vec{w}_M=\vec{w}_M(t,x)=(h(t,x),h(t,x)u_m(t,x),h(t,x)\alpha_1(t,x),\ldots,h(t,x)\alpha_M(t,x))^T\in\mathbb{R}^{M+2},
\end{equation} 
transport system matrix 
\begin{equation}\label{eq:transport-matrix-definition}
	A_{M}(\vec{w}_M):=\frac{\partial \vec{F}_M(\vec{w}_M)}{\partial \vec{w}_M}-Q_M(\vec{w}_M)\in\mathbb{R}^{(M+2)\times(M+2)},
\end{equation} 
and source term \(\vec{S}_M(\vec{w}_M)\in\mathbb{R}^{M+2}\). Note that the non-conservative flux matrix \(Q_M(\vec{w}_M)\) is defined by the relation \(\vec{Q}_M(\vec{w}_M)=Q_M(\vec{w}_M)\partial_x \vec{w}_M\).

This non-conservative system of partial differential equations \eqref{eq:SWME-compact} is only hyperbolic for \(M=0\) and \(M=1\) \cite{SWME}. This motivated the authors of \cite{HSWME} to regularize the system matrix such that it has real eigenvalues, obtaining the so-called Hyperbolic Shallow Water Moment Equations (HSWME), which are hyperbolic for arbitrary order \(M\). Recently, the HSWME have been extended to axisymmetric flow \cite{ASWME_Verbiest} and 2D flow \cite{Bauerle_Rotational}. Other variations include SWME models for bedload problems \cite{SWME_bedload}, dispersive SWME \cite{scholz_dispersion_2023}, and SWME with the Chézy friction model \cite{Steldermann_2023}. In this paper, the standard SWME as derived in \cite{SWME} are used for the analysis of the adaptive procedure and the numerical simulations. However, the adaptive procedure in this paper can be applied to each of the SWME models mentioned in this section with only minor adjustments.

\begin{remark}
    Although a spatially varying bottom topography $h_b(x)$ would be of interest, its inclusion would require a well-balanced treatment, which is beyond the scope of this paper and is left for future work. For strongly varying bottom topography, it may furthermore be beneficial to incorporate the topographic gradient into the domain decomposition criteria proposed in Section \ref{sec:domain-decomposition}, since large flow gradients are expected to develop over regions of varying depth.
\end{remark}


\paragraph{Notation} In the remainder of this paper, we will denote a SWME model of a certain order \(M\) by the short notation \(\text{SWME}_{M}\).

\subsection{Examples}
In this section, we will briefly state three specific examples of the SWME: the \(\text{SWME}_{0}\) (\(M=0\)), the \(\text{SWME}_{1}\) (\(M=1\)) and the \(\text{SWME}_{2}\) (\(M=2\)), first derived in \cite{SWME}. These examples will be used throughout the paper to illustrate the error estimation and domain decomposition in Section \ref{sec:domain-decomposition}. The numerical simulations in Section \ref{section:Numerics}, however, will be run with larger orders \(M\).

\paragraph{\(\text{SWME}_{0}\)}
For \(M=0\), the SWME \eqref{eq:SWME-h}-\eqref{eq:SWME-halphai} reduce to the SWE and read
\begin{equation}\label{SWME-order0}
	\partial_t
	\begin{pmatrix}
		h \\[5pt]
		hu_m
	\end{pmatrix}
	+ 
	\begin{pmatrix}
		0 & 1 \\[5pt]
		gh - u_m^2 & 2u_m
	\end{pmatrix}
	\partial_x
	\begin{pmatrix}
		h \\[5pt]
		hu_m
	\end{pmatrix}
	= -\frac{\nu}{\lambda}
	\begin{pmatrix}
		0 \\[5pt]
		u_m
	\end{pmatrix}.
\end{equation}
The velocity is constant in the vertical variable: \(u(t,x,\zeta)=u_m(t,x)\).

\paragraph{\(\text{SWME}_{1}\)}
The \(\text{SWME}_{1}\) are given by \eqref{eq:SWME-h}-\eqref{eq:SWME-halphai} with \(M=1\) and read
\begin{equation}\label{SWME-order1}
	\partial_t
	\begin{pmatrix}
		h \\[5pt]
		hu_m \\[5pt]
		h\alpha_1
	\end{pmatrix}
	+ 
	\begin{pmatrix}
		0 & 1 & 0 \\[5pt]
		gh - u_m^2 - \frac{\alpha_1^2}{3} & 2u_m & \frac{2\alpha_1}{3} \\[5pt]
		-2u_m\alpha_1 & 2\alpha_1 & u_m 
	\end{pmatrix}
	\partial_x
	\begin{pmatrix}
		h \\[5pt]
		hu_m \\[5pt]
		h\alpha_1
	\end{pmatrix}
	= -\frac{\nu}{\lambda}
	\begin{pmatrix}
		0 \\[5pt]
		u_m + \alpha_1 \\[5pt]
		3\left(u_m+\alpha_1+4\frac{\lambda}{h}\alpha_1\right)
	\end{pmatrix}.
\end{equation}
The velocity is linear in the vertical variable: \(u(t,x,\zeta)=u_m(t,x)+\alpha_1(t,x)\phi_1(\zeta)\).

\paragraph{\(\text{SWME}_{2}\)}
The \(\text{SWME}_{2}\) are given by \eqref{eq:SWME-h}-\eqref{eq:SWME-halphai} with \(M=2\) and read
\begin{equation}\label{SWME-order2}
	\partial_t
	\begin{pmatrix}
		h \\[5pt]
		hu_m \\[5pt]
		h\alpha_1 \\[5pt]
		h\alpha_2
	\end{pmatrix}
	+
	\begin{pmatrix}
		0 & 1 & 0 & 0 \\[5pt]
		gh - u_m^2 - \frac{\alpha_1^2}{3} - \frac{\alpha_2^2}{5} & 2u_m &\frac{2}{3}\alpha_1 & \frac{2}{5}\alpha_2 \\[5pt]
		-2u_m\alpha_1 - \frac{4}{5}\alpha_1\alpha_2 & 2\alpha_1 & u_m +\alpha_2 & \frac{3}{5}\alpha_1 \\[5pt]
		-\frac{2}{3}\alpha_1^2-2\alpha_2u_m-\frac{2}{7}\alpha_2^2 & 2\alpha_2 & \frac{1}{3}\alpha_1 & u_m +\frac{3}{7}\alpha_2
	\end{pmatrix}
	\partial_x
	\begin{pmatrix}
		h \\[5pt]
		hu_m \\[5pt]
		h\alpha_1 \\[5pt]
		h\alpha_2
	\end{pmatrix}
	= -\frac{\nu}{\lambda}
	\begin{pmatrix}
		0 \\[5pt]
		u_m + \alpha_1 +\alpha_2 \\[5pt]
		3\left(u_m+\alpha_1+\alpha_2+4\frac{\lambda}{h}\alpha_1\right) \\[5pt]
		5\left(u_m+\alpha_1+\alpha_2+12\frac{\lambda}{h}\alpha_2\right)
	\end{pmatrix}.
\end{equation}
The velocity is quadratic in the vertical variable: \(u(t,x,\zeta)=u_m(t,x)+\alpha_1(t,x)\phi_1(\zeta)+\alpha_2(t,x)\phi_2(\zeta)\).

\section{Adaptive Simulation of Shallow Water Moment Equations}\label{section:Introduction_SpatiallyAdaptiveSimulation}
In this section, an algorithm for the model-adaptive simulation of the SWME will be constructed. The algorithm consists of two steps: (1) the first step, discussed in Section \ref{sec:domain-decomposition}, is a domain decomposition, dividing the domain into subdomains, each modeled by a SWME model of an appropriate order \(M\), where the appropriate order is determined by a set of error estimators that estimate the error made by restricting the velocity expansion \eqref{eq:legendre-sum} to the first \(M\) moments and Legendre polynomials. (2) The second step, discussed in Section \ref{sec:spatial_coupling}, is the spatial coupling of the subdomains at the boundary interfaces of the subdomains, i.e., the interfaces that separate two subdomains modeled by SWME models with different orders. 

\paragraph{Spatially- and time-adaptive order}
So far, the order \(M\) of the SWME model has been treated as fixed and constant in time and space throughout the literature. As argued in the introduction, in many applications, some subdomains require more modeling complexity than other subdomains. This motivates the need for a spatially-adaptive order. Moreover, the number of subdomains and the boundaries of the subdomains change in time as the waves travel through the fluid domain. This motivates the need for a time-adaptive order. Instead of assuming a constant order \(M\), we will therefore consider a space- and time-dependent order \(M= M(t,x)\) in this paper.

\subsection{Domain decomposition}\label{sec:domain-decomposition}
In this paper, we propose error estimators based on a decomposition of a higher-order \(\text{SWME}_{M_\mathrm{H}}\), with \(M_\mathrm{H}\in\mathbb{N}\). The idea is to use the hierarchical structure of the SWME, so that the effect of adding higher-order moments can be directly estimated. This allows for an efficient model error approximation. Using the hierarchical structure of the equations \eqref{eq:SWME-compact}, a lower-order \(\text{SWME}_{M_\mathrm{L}}\), with \(M_\mathrm{L}\in\mathbb{N}\) and where \(M_\mathrm{L}<M_\mathrm{H}\), can be identified within the decomposition of the higher-order \(\text{SWME}_{M_\mathrm{H}}\), such that the deviation of the higher-order \(\text{SWME}_{M_\mathrm{H}}\) from the lower-order \(\text{SWME}_{M_\mathrm{L}}\) can be evaluated. The error estimators are then used to decompose the domain into subdomains, each modeled by a SWME model of an appropriate order. This is done by evaluating whether the error estimators exceed chosen thresholds, which results in a set of domain decomposition criteria for model-coarsening, i.e., reducing the order of the SWME model from \(M_\mathrm{H}\) to \(M_\mathrm{L}\), and for model-refinement, i.e., increasing the order of the SWME model from \(M_\mathrm{L}\) to \(M_\mathrm{H}\). The analytical form of the domain decomposition criteria for model-coarsening and model-refinement are given in Definition \ref{def:coarsening} and Definition \ref{def:refining}, respectively. For numerical simulations, the analytical domain decomposition criteria are numerically evaluated on the grid. This is discussed in Section \ref{section:numerical-evaluation-criteria}, in which the numerical evaluation of the analytical domain decomposition criteria is summarized in expressions \eqref{eq:res1_approximated}, \eqref{eq:res2_approximated} and \eqref{eq:increase-criteria-physical_discrete}.

\begin{remark}
	Many existing variants of shallow water moment models have a hierarchical structure as well, not only the SWME derived in \cite{SWME}, such that the same ideas can be applied accordingly.
\end{remark}

Let us first introduce the following notation. The higher-order state variable vector \(\vec{w}_{M_\mathrm{H}}\in\mathbb{R}^{M_\mathrm{H}+2}\) \eqref{eq:state-variable-vector} is decomposed as
\begin{equation}\label{eq:state-variable-vector_decomposition}
	\vec{w}_{M_\mathrm{H}} =
	\begin{pmatrix}
		\vec{w}_{:M_\mathrm{L}} \\[5pt]
		\vec{w}_{M_\mathrm{L}:M_\mathrm{H}}
	\end{pmatrix},
\end{equation}
where \(\vec{w}_{:M_\mathrm{L}}\in\mathbb{R}^{M_\mathrm{L}+2}\) contains the first \(M_\mathrm{L}+2\) variables of \(\vec{w}_{M_\mathrm{H}}\) and where \(\vec{w}_{M_\mathrm{L}:M_\mathrm{H}}\in\mathbb{R}^{M_\mathrm{H}-M_\mathrm{L}}\) contains the remaining variables.

The system matrix \(A_{M_\mathrm{H}}(\vec{w}_{M_\mathrm{H}})\in\mathbb{R}^{(M_\mathrm{H}+2)\times(M_\mathrm{H}+2)}\) of the higher-order \(\text{SWME}_{M_\mathrm{H}}\) \eqref{eq:SWME-compact} is accordingly decomposed as 
\begin{equation}\label{eq:SWME-MHth_order}
	A_{M_\mathrm{H}}(\vec{w}_{M_\mathrm{H}}) =
	\begin{pmatrix}
		A_{:M_\mathrm{L},:M_\mathrm{L}}(\vec{w}_{M_\mathrm{H}}) & A_{:M_\mathrm{L},M_\mathrm{L}:M_\mathrm{H}}(\vec{w}_{M_\mathrm{H}}) \\[5pt]
		A_{M_\mathrm{L}:M_\mathrm{H},:M_\mathrm{L}}(\vec{w}_{M_\mathrm{H}}) & A_{M_\mathrm{L}:M_\mathrm{H},M_\mathrm{L}:M_\mathrm{H}}(\vec{w}_{M_\mathrm{H}})
	\end{pmatrix},
\end{equation}
where:
\begin{itemize}
	\item \(A_{:M_\mathrm{L},:M_\mathrm{L}}(\vec{w}_{M_\mathrm{H}})\in\mathbb{R}^{(M_\mathrm{L}+2)\times (M_\mathrm{L}+2)}\) are the entries in the first \(M_\mathrm{L}+2\) rows and the first \(M_\mathrm{L}+2\) columns of \(A_{M_\mathrm{H}}(\vec{w}_{M_\mathrm{H}})\), 
	\item \(A_{:M_\mathrm{L},M_\mathrm{L}:M_\mathrm{H}}(\vec{w}_{M_\mathrm{H}})\in\mathbb{R}^{(M_\mathrm{L}+2)\times (M_\mathrm{H}-M_\mathrm{L})}\) are the entries in the first \(M_\mathrm{L}+2\) rows and the last \(M_\mathrm{H}-M_\mathrm{L}\) columns of \(A_{M_\mathrm{H}}(\vec{w}_{M_\mathrm{H}})\), 
	\item \(A_{M_\mathrm{L}:M_\mathrm{H},:M_\mathrm{L}}(\vec{w}_{M_\mathrm{H}})\in\mathbb{R}^{ (M_\mathrm{H}-M_\mathrm{L})\times (M_\mathrm{L}+2)}\) are the entries in the first \(M_\mathrm{L}+2\) columns and the last \(M_\mathrm{H}-M_\mathrm{L}\) rows of \(A_{M_\mathrm{H}}(\vec{w}_{M_\mathrm{H}})\),
	\item \(A_{M_\mathrm{L}:M_\mathrm{H},M_\mathrm{L}:M_\mathrm{H}}(\vec{w}_{M_\mathrm{H}})\in\mathbb{R}^{(M_\mathrm{H}-M_\mathrm{L})\times (M_\mathrm{H}-M_\mathrm{L})}\) are the entries in the last \(M_\mathrm{H}-M_\mathrm{L}\) rows and the last \(M_\mathrm{H}-M_\mathrm{L}\) columns of \(A_{M_\mathrm{H}}(\vec{w}_{M_\mathrm{H}})\). 
\end{itemize}
Finally, the source term \(\vec{S}_{M_\mathrm{H}}(\vec{w}_{M_\mathrm{H}})\in\mathbb{R}^{M_\mathrm{H}+2}\) of the \(\text{SWME}_{M_\mathrm{H}}\) is decomposed as 
\begin{equation}\label{eq:decomposition-sourceTerm}
	\vec{S}_{M_\mathrm{H}}(\vec{w}_{M_\mathrm{H}}) =
	\begin{pmatrix}
		\vec{S}_{:M_\mathrm{L}}(\vec{w}_{M_\mathrm{H}}) \\
		\vec{S}_{M_\mathrm{L}:M_\mathrm{H}}(\vec{w}_{M_\mathrm{H}})
	\end{pmatrix},
\end{equation}
where \(\vec{S}_{:M_\mathrm{L}}(\vec{w}_{M_\mathrm{H}})\in\mathbb{R}^{M_\mathrm{L}+2}\) contains the first \(M_\mathrm{L}+2\) entries of \(\vec{S}_{M_\mathrm{H}}(\vec{w}_{M_\mathrm{H}})\) and where \(\vec{S}_{M_\mathrm{L}:M_\mathrm{H}}(\vec{w}_{M_\mathrm{H}})\in\mathbb{R}^{M_\mathrm{H}-M_\mathrm{L}}\) contains the remaining entries.
Inserting the decompositions \eqref{eq:SWME-MHth_order} and \eqref{eq:decomposition-sourceTerm} in the general form \eqref{eq:SWME-compact}, the \(\text{SWME}_{M_\mathrm{H}}\) \eqref{eq:SWME-compact} can be written as
\begin{equation}\label{eq:SWME-decomposed}
	\partial_t
	\begin{pmatrix}
		\vec{w}_{:M_\mathrm{L}} \\[5pt]
		\vec{w}_{M_\mathrm{L}:M_\mathrm{H}} 
	\end{pmatrix}
	+
	\begin{pmatrix}
		A_{:M_\mathrm{L},:M_\mathrm{L}}(\vec{w}_{M_\mathrm{H}}) & A_{:M_\mathrm{L},M_\mathrm{L}:M_\mathrm{H}}(\vec{w}_{M_\mathrm{H}}) \\[5pt]
		A_{M_\mathrm{L}:M_\mathrm{H},:M_\mathrm{L}}(\vec{w}_{M_\mathrm{H}}) & A_{M_\mathrm{L}:M_\mathrm{H},M_\mathrm{L}:M_\mathrm{H}}(\vec{w}_{M_\mathrm{H}})
	\end{pmatrix}
	\partial_x
	\begin{pmatrix}
		\vec{w}_{:M_\mathrm{L}} \\[5pt]
		\vec{w}_{M_\mathrm{L}:M_\mathrm{H}} 
	\end{pmatrix} = 
	\begin{pmatrix}
		\vec{S}_{:M_\mathrm{L}}(\vec{w}_{M_\mathrm{H}}) \\
		\vec{S}_{M_\mathrm{L}:M_\mathrm{H}}(\vec{w}_{M_\mathrm{H}})
	\end{pmatrix},
\end{equation}
while the \(\text{SWME}_{M_\mathrm{L}}\) \eqref{eq:SWME-compact} reads
\begin{equation}\label{eq:SWME-Ml}
	\partial_t \vec{w}_{M_\mathrm{L}} + A_{M_\mathrm{L}}(\vec{w}_{M_\mathrm{L}})\partial_x \vec{w}_{M_\mathrm{L}} = \vec{S}_{M_\mathrm{L}}(\vec{w}_{M_\mathrm{L}}),
\end{equation}
with system matrix \(A_{M_\mathrm{L}}(\vec{w}_{M_\mathrm{L}})\in\mathbb{R}^{(M_\mathrm{L}+2)\times(M_\mathrm{L}+2)}\) and source term \(\vec{S}_{M_\mathrm{L}}(\vec{w}_{M_\mathrm{L}})\in\mathbb{R}^{M_\mathrm{L}+2}\).

From the hierarchical structure of the SWME \eqref{eq:SWME-h}-\eqref{eq:SWME-halphai} (lower-order SWME models are included as a special case in higher-order SWME models), it follows that \(A_{:M_\mathrm{L},:M_\mathrm{L}}(\vec{w}_{M_\mathrm{H}})\) and \(\vec{S}_{:M_\mathrm{L}}(\vec{w}_{M_\mathrm{H}})\) can be written as
\begin{align}\label{eq:A-S-decomposition1}
	&A_{:M_\mathrm{L},:M_\mathrm{L}}(\vec{w}_{M_\mathrm{H}}) = A_{M_\mathrm{L}}(\vec{w}_{:M_\mathrm{L}})+\delta A_{M_\mathrm{L},M_\mathrm{H}}(\vec{w}_{M_\mathrm{H}}), \\[5pt] \label{eq:A-S-decomposition2}&\vec{S}_{:M_\mathrm{L}}(\vec{w}_{M_\mathrm{H}})= \vec{S}_{M_\mathrm{L}}(\vec{w}_{:M_\mathrm{L}}) + \delta \vec{S}_{M_\mathrm{L},M_\mathrm{H}}(\vec{w}_{M_\mathrm{H}}),
\end{align}
where \(\delta A_{M_\mathrm{L},M_\mathrm{H}}(\vec{w}_{M_\mathrm{H}})\in\mathbb{R}^{(M_\mathrm{L}+2)\times(M_\mathrm{L}+2)}\) represents the difference between the system matrices of the \(\text{SWME}_{M_\mathrm{L}}\) and the \(\text{SWME}_{M_\mathrm{H}}\) for the first \(M_\mathrm{L}+2\) variables and where \(\delta \vec{S}_{M_\mathrm{L},M_\mathrm{H}}(\vec{w}_{M_\mathrm{H}})\in\mathbb{R}^{M_\mathrm{L}+2}\) represents the difference between the source terms of the \(\text{SWME}_{M_\mathrm{L}}\) and the \(\text{SWME}_{M_\mathrm{H}}\) for the first \(M_\mathrm{L}+2\) variables. The analytical formulas for \(\delta A_{M_\mathrm{L},M_\mathrm{H}}(\vec{w}_{M_\mathrm{H}})\) and \(\delta \vec{S}_{M_\mathrm{L},M_\mathrm{H}}(\vec{w}_{M_\mathrm{H}})\) are given in \ref{sec:appendix}. Comparing the \(\text{SWME}_{M_\mathrm{H}}\) \eqref{eq:SWME-decomposed} with the \(\text{SWME}_{M_\mathrm{L}}\) \eqref{eq:SWME-Ml}, one can see that if the equation
\begin{equation}\label{eq:domain-decomposition-criterion-1}
	\begin{pmatrix}
		\delta A_{M_\mathrm{L},M_\mathrm{H}}(\vec{w}_{M_\mathrm{H}}(t^*,x^*)) \\[5pt]
		A_{:M_\mathrm{L},M_\mathrm{L}:M_\mathrm{H}}(\vec{w}_{M_\mathrm{H}}(t^*,x^*))
	\end{pmatrix}^T
	\partial_x
	\begin{pmatrix}
		\vec{w}_{:M_\mathrm{L}}(t^*,x^*)\\[5pt]
		\vec{w}_{M_\mathrm{L}:M_\mathrm{H}}(t^*,x^*)
	\end{pmatrix}
	=\delta \vec{S}_{M_\mathrm{L},M_\mathrm{H}}(\vec{w}_{M_\mathrm{H}}(t^*,x^*))
\end{equation}
holds for some \(t^*\in[0,T]\) and some \(x^*\in\Omega_x\), then the \(\text{SWME}_{M_\mathrm{H}}\) \eqref{eq:SWME-decomposed} is equivalent to the \(\text{SWME}_{M_\mathrm{L}}\) \eqref{eq:SWME-Ml} for the first \(M_\mathrm{L}+2\) variables, meaning that the high-order moments $\vec{w}_{M_L:M_H}$ have no influence on the equations for the lower-order moments $\vec{w}_{:M_L}$, at \((t^*,x^*)\). Note that if the higher-order variables \(\vec{w}_{M_\mathrm{L}:M_\mathrm{H}}(t^*,x^*)=\vec{0}\in\mathbb{R}^{M_\mathrm{H}-M_\mathrm{L}}\), then it follows from the hierarchical structure of the SWME \eqref{eq:SWME-h}-\eqref{eq:SWME-halphai} and the derivation in \ref{sec:appendix} that 
\begin{align}\label{eq:SWME_diff_SMh_zeroLastMoment}
	&\vec{S}_{:M_\mathrm{L}}(\vec{w}^{(0)}_{M_\mathrm{H}}(t^*,x^*))=\vec{S}_{M_\mathrm{L}}(\vec{w}_{:M_\mathrm{L}}(t^*,x^*)), \\[5pt]\label{eq:SWME_diff_AMh_zeroLastMoment}
	&A_{:M_\mathrm{L},:M_\mathrm{L}}(\vec{w}^{(0)}_{M_\mathrm{H}}(t^*,x^*)) = A_{M_\mathrm{L}}(\vec{w}_{:M_\mathrm{L}}(t^*,x^*)),
\end{align} 
where we defined \(\vec{w}^{(0)}_{M_\mathrm{H}}(t^*,x^*):=\left(\vec{w}_{:M_\mathrm{L}}(t^*,x^*)^T,\vec{0}^T\right)^T\in\mathbb{R}^{M_\mathrm{H}+2}\). Thus, under the assumption that \(|\alpha_i(t^*,x^*)|\ll 1\) for \(i=M_\mathrm{L}+1,\ldots,M_\mathrm{H}\), then \(\delta A_{M_\mathrm{L},M_\mathrm{H}}(\vec{w}_{M_\mathrm{H}}(t^*,x^*))\) can be neglected and Equation \eqref{eq:domain-decomposition-criterion-1} reduces to
\begin{equation}\label{eq:domain-decomposition-criterion-1_reduced}
	A_{:M_\mathrm{L},M_\mathrm{L}:M_\mathrm{H}}(\vec{w}_{M_\mathrm{H}}(t^*,x^*))\partial_x\vec{w}_{M_\mathrm{L}:M_\mathrm{H}}(t^*,x^*)=\vec{0}.
\end{equation}
We further require that \(\partial_t\vec{w}_{M_\mathrm{L}:M_\mathrm{H}}(t^*,x^*)=\vec{0}\in\mathbb{R}^{M_\mathrm{H}-M_\mathrm{L}}\), which holds if the equation
\begin{equation}\label{eq:domain-decomposition-criterion-2}
	\begin{pmatrix}
		A_{M_\mathrm{L}:M_\mathrm{H},:M_\mathrm{L}}(\vec{w}_{M_\mathrm{H}}(t^*,x^*)) \\[5pt] A_{M_\mathrm{L}:M_\mathrm{H},M_\mathrm{L}:M_\mathrm{H}}(\vec{w}_{M_\mathrm{H}}(t^*,x^*))
	\end{pmatrix}^T\partial_x
	\begin{pmatrix}
		\vec{w}_{:M_\mathrm{L}}(t^*,x^*) \\[5pt]
		\vec{w}_{M_\mathrm{L}:M_\mathrm{H}}(t^*,x^*) 
	\end{pmatrix} = 
	\vec{S}_{M_\mathrm{L}:M_\mathrm{H}}(\vec{w}_{M_\mathrm{H}}(t^*,x^*))
\end{equation}
is satisfied at \((t^*,x^*)\). Equation \eqref{eq:domain-decomposition-criterion-2} ensures that the assumption \(|\alpha_i(t^*,x^*)|\ll 1\), \(i=M_\mathrm{L}+1,\ldots,M_\mathrm{H}\), and Equation \eqref{eq:domain-decomposition-criterion-1_reduced} remain valid around \(t^*\). Equations \eqref{eq:domain-decomposition-criterion-1_reduced} and \eqref{eq:domain-decomposition-criterion-2} are fulfilled if the higher-order \(\text{SWME}_{M_\mathrm{H}}\) model does not add significant high-order moment contributions and the lower-order \(\text{SWME}_{M_\mathrm{L}}\) model is sufficient.

\paragraph{Model-coarsening criteria}
Assume that the flow is modeled by the higher-order \(\text{SWME}_{M_\mathrm{H}}\) model. To identify where the order of the SWME model can be reduced from higher order \(M_\mathrm{H}\) to lower order \(M_\mathrm{L}\) without losing much modeling complexity, the validity of Equations \eqref{eq:domain-decomposition-criterion-1_reduced} and \eqref{eq:domain-decomposition-criterion-2} is verified using their respective numerical residuals. Define the residuals of Equations \eqref{eq:domain-decomposition-criterion-1_reduced} and \eqref{eq:domain-decomposition-criterion-2} at \((t^*,x^*)\) by \(\mathcal{R}_{:M_\mathrm{L}}(t^*,x^*)\in\mathbb{R}^{M_\mathrm{L}+2}\) and \(\mathcal{R}_{M_\mathrm{L}:M_\mathrm{H}}(t^*,x^*)\in\mathbb{R}^{M_\mathrm{H}-M_\mathrm{L}}\), respectively:
\begin{align}
	\label{eq:res1}
	&\mathcal{R}_{:M_\mathrm{L}}(t^*,x^*):=A_{:M_\mathrm{L},M_\mathrm{L}:M_\mathrm{H}}(\vec{w}_{M_\mathrm{H}}(t^*,x^*))\partial_x\vec{w}_{M_\mathrm{L}:M_\mathrm{H}}(t^*,x^*),\\[5pt]\label{eq:res2}
	&\mathcal{R}_{M_\mathrm{L}:M_\mathrm{H}}(t^*,x^*):= 
	\begin{pmatrix}
		A_{M_\mathrm{L}:M_\mathrm{H},:M_\mathrm{L}}(\vec{w}_{M_\mathrm{H}}(t^*,x^*)) \\[5pt] A_{M_\mathrm{L}:M_\mathrm{H},M_\mathrm{L}:M_\mathrm{H}}(\vec{w}_{M_\mathrm{H}}(t^*,x^*))
	\end{pmatrix}^T\partial_x
	\begin{pmatrix}
		\vec{w}_{:M_\mathrm{L}}(t^*,x^*) \\[5pt]
		\vec{w}_{M_\mathrm{L}:M_\mathrm{H}}(t^*,x^*) 
	\end{pmatrix} - 
	\vec{S}_{M_\mathrm{L}:M_\mathrm{H}}(\vec{w}_{M_\mathrm{H}}(t^*,x^*)).
\end{align}
The norm of the residual \(\mathcal{R}_{:M_\mathrm{L}}(t^*,x^*)\) determines how large the deviations corresponding to the gradients of the higher-order variables \(\vec{w}_{M_\mathrm{L}:M_\mathrm{H}}(t^*,x^*)\) are, while the norm of the residual \(\mathcal{R}_{M_\mathrm{L}:M_\mathrm{H}}(t^*,x^*)\) quantifies the change of the higher-order variables \(\vec{w}_{M_\mathrm{L}:M_\mathrm{H}}(t^*,x^*)\) in time. The two residuals \(\mathcal{R}_{:M_\mathrm{L}}(t^*,x^*)\) \eqref{eq:res1} and \(\mathcal{R}_{M_\mathrm{L}:M_\mathrm{H}}(t^*,x^*)\) \eqref{eq:res2} can be interpreted as error estimators. Then, these error estimators are used to construct model-coarsening criteria that identify regions where the order can be reduced. 
\begin{definition}[Model-coarsening criteria]\label{def:coarsening}
	Let \((t^*,x^*)\in[0,T]\times \Omega_x\) and let \(M_\mathrm{H},M_\mathrm{L}\in\mathbb{N}\), with \(M_\mathrm{L}<M_\mathrm{H}\). Define the residuals \(\mathcal{R}_{:M_\mathrm{L}}(t^*,x^*)\) and \(\mathcal{R}_{M_\mathrm{L}:M_\mathrm{H}}(t^*,x^*)\) by \eqref{eq:res1} and \eqref{eq:res2}, respectively. The set of model-coarsening criteria at \((t^*,x^*)\) is given by  
	\begin{equation}\label{eq:residual_decrease_criteria}
		\left\Vert\vec{w}_{M_\mathrm{L}:M_\mathrm{H}}(t^*,x^*)\right\Vert_{\infty}<\epsilon_w, \quad
		\left\Vert\mathcal{R}_{:M_\mathrm{L}}(t^*,x^*)\right\Vert_2 < \epsilon_{\mathcal{R}_{:M_\mathrm{L}}},\quad \left\Vert\mathcal{R}_{M_\mathrm{L}:M_\mathrm{H}}(t^*,x^*)\right\Vert_2 < \epsilon_{\mathcal{R}_{M_\mathrm{L}:M_\mathrm{H}}}.
	\end{equation}
	If all three model-coarsening criteria in \eqref{eq:residual_decrease_criteria} are satisfied, then the order is reduced from higher order \(M_\mathrm{H}\) to lower order \(M_\mathrm{L}\), where \(\epsilon_w\), \(\epsilon_{\mathcal{R}_{:M_\mathrm{L}}},\epsilon_{\mathcal{R}_{M_\mathrm{L}:M_\mathrm{H}}}\in\mathbb{R}^+\) are chosen tolerances. 
\end{definition} 
The choice of the max norm \(||\cdot||_{\infty}\) in \(\left\Vert\vec{w}_{M_\mathrm{L}:M_\mathrm{H}}(t^*,x^*)\right\Vert_{\infty}<\epsilon_w\) is motivated by the fact that the magnitude of each of the higher-order variables in \(\vec{w}_{M_\mathrm{L}:M_\mathrm{H}}(t^*,x^*)\) needs to be small for Equations \eqref{eq:SWME_diff_SMh_zeroLastMoment} and \eqref{eq:SWME_diff_AMh_zeroLastMoment} to hold approximately.

\paragraph{Model-refinement criteria} Assume now that the flow is modeled by the lower-order \(\text{SWME}_{M_\mathrm{L}}\) model. To identify where the order of the SWME model should be increased from lower order \(M_\mathrm{L}\) to higher order \(M_\mathrm{H}\) to capture the complex fluid flow dynamics, the following rationale will be used. Note that the systems \eqref{eq:SWME-decomposed} and \eqref{eq:SWME-Ml} cannot be compared in this case, as the \(\text{SWME}_{M_\mathrm{L}}\) model does not include the higher-order variables \(\vec{w}_{M_\mathrm{L}:M_\mathrm{H}}\). To circumvent this problem, we assume that the \(\text{SWME}_{M_\mathrm{L}}\) and the \(\text{SWME}_{M_\mathrm{H}}\) are equivalent at \((t^*,x^*)\) in the specific case where 
\begin{equation}\label{eq:increase-assumptions}
	\vec{w}_{M_\mathrm{L}:M_\mathrm{H}}(t^*,x^*)=\vec{0}\in\mathbb{R}^{M_\mathrm{H}-M_\mathrm{L}} \qquad \text{and} \qquad \partial_x\vec{w}_{M_\mathrm{L}:M_\mathrm{H}}(t^*,x^*)=\vec{0}\in\mathbb{R}^{M_\mathrm{H}-M_\mathrm{L}}.
\end{equation}
Note that the higher-order variables \(\vec{w}_{M_\mathrm{L}:M_\mathrm{H}}(t^*,x^*)\) are not even defined at \((t^*,x^*)\), but that we assume that the \(\text{SWME}_{M_\mathrm{L}}\) is actually originating from the \(\text{SWME}_{M_\mathrm{H}}\) with the higher-order variables equal to zero. Under this assumption, Equation \eqref{eq:domain-decomposition-criterion-2} reduces to
\begin{equation}\label{eq:domain-decomposition-criterion-increasing}
	A_{M_\mathrm{L}:M_\mathrm{H},:M_\mathrm{L}}(\vec{w}^{(0)}_{M_\mathrm{H}}(t^*,x^*))\partial_x\vec{w}_{:M_\mathrm{L}}(t^*,x^*)=
	\vec{S}_{M_\mathrm{L}:M_\mathrm{H}}(\vec{w}^{(0)}_{M_\mathrm{H}}(t^*,x^*)).
\end{equation}
One can easily verify that under the assumptions \eqref{eq:increase-assumptions} and \eqref{eq:domain-decomposition-criterion-increasing}, the \(\text{SWME}_{M_\mathrm{L}}\) is indeed equivalent to the \(\text{SWME}_{M_\mathrm{H}}\). For the equivalence between the lower order \(\text{SWME}_{M_\mathrm{L}}\) and the higher order \(\text{SWME}_{M_\mathrm{H}}\) to hold in time around \(t=t^*\), Equation \eqref{eq:domain-decomposition-criterion-increasing} needs to be satisfied. Analogously to the identification of regions where the order can be reduced from higher-order \(M_\mathrm{H}\) to lower-order \(M_\mathrm{L}\) using equations \eqref{eq:domain-decomposition-criterion-1_reduced} and \eqref{eq:domain-decomposition-criterion-2}, one could calculate the magnitude of the residual of Equation \eqref{eq:domain-decomposition-criterion-increasing}. However, the quantity \(A_{M_\mathrm{L}:M_\mathrm{H},:M_\mathrm{L}}(\vec{w}^{(0)}_{M_\mathrm{H}}(t^*,x^*))\partial_x\vec{w}_{:M_\mathrm{L}}(t^*,x^*)\) appearing in Equation \eqref{eq:domain-decomposition-criterion-increasing} cannot be calculated since the flow at \((t^*,x^*)\) is modeled by the lower-order \(\text{SWME}_{M_\mathrm{L}}\). Calculating the residual of Equation \eqref{eq:domain-decomposition-criterion-increasing} would therefore require solving additional equations. To avoid additional calculations, we will therefore
verify whether a particular solution to \eqref{eq:domain-decomposition-criterion-increasing} that is easy to calculate is satisfied. One particular solution to \eqref{eq:domain-decomposition-criterion-increasing} at \((t^*,x^*)\) that is easy to verify is given by the equalities \eqref{eq:increase-assumptions} and the additional equalities
\begin{equation}\label{eq:increase-relations}
	\vec{S}_{M_\mathrm{L}:M_\mathrm{H}}(\vec{w}^{(0)}_{M_\mathrm{H}}(t^*,x^*))=\vec{0} \in \mathbb{R}^{M_\mathrm{H}-M_\mathrm{L}}\qquad \text{and} \qquad \partial_x\vec{w}_{:M_\mathrm{L}}(t^*,x^*) = \vec{0}\in\mathbb{R}^{M_\mathrm{L}}.
\end{equation}
The left-hand sides of the expressions in \eqref{eq:increase-relations} can be interpreted as error estimators for refining the model, and both describe the contribution of the lower-order moments $\vec{w}_{:M_\mathrm{L}}(t^*,x^*)$ to the equations for the higher-order moments $\vec{w}_{M_\mathrm{L}:M_\mathrm{R}}(t^*,x^*)$. While the left-hand side of the first equality in \eqref{eq:increase-relations} describes the contribution of $\vec{w}_{:M_\mathrm{L}}(t^*,x^*)$ to the friction terms in the equations for $\vec{w}_{M_\mathrm{L}:M_\mathrm{R}}(t^*,x^*)$, the left-hand side of the second equality in \eqref{eq:increase-relations} measures the spatial gradients of $\vec{w}_{:M_\mathrm{L}}(t^*,x^*)$ that contribute to the transport part in the equations for $\vec{w}_{M_\mathrm{L}:M_\mathrm{R}}(t^*,x^*)$. Then, these error estimators are used to construct model-refinement criteria that identify regions where the order should be increased. 
\begin{definition}[Model-refinement criteria]\label{def:refining}
	Let \((t^*,x^*)\in[0,T]\times \Omega_x\) and let \(M_\mathrm{H},M_\mathrm{L}\in\mathbb{N}\), with \(M_\mathrm{L}<M_\mathrm{H}\). The set of model-refinement criteria is given by  
	\begin{equation}\label{eq:increase-criteria-physical}
		\left\Vert\vec{S}_{M_\mathrm{L}:M_\mathrm{H}}(\vec{w}^{(0)}_{M_\mathrm{H}}(t^*,x^*))\right\Vert_{\infty}>\kappa_S,\quad \left\Vert\partial_x\vec{w}_{:M_\mathrm{L}}(t^*,x^*)\right\Vert_{\infty}>\kappa_{\partial}.
	\end{equation}
	The order is increased if one of the model-refinement criteria in \eqref{eq:increase-criteria-physical} is satisfied, where \(\kappa_S\in\mathbb{R}^+\) and \(\kappa_{\partial}\in\mathbb{R}^+\) are chosen tolerances.
\end{definition} 
\begin{remark}
	The domain decomposition criteria proposed in definitions \ref{def:coarsening} and \ref{def:refining} can be viewed as an extension of classical AMR mesh-refinement criteria related to the gradient of the height and the gradient of the momentum, which are included in the second criterion in \eqref{eq:increase-criteria-physical} in Definition \ref{def:refining}, for the shallow water equations \cite{bader_dynamically_2010,AAMM-12-2}. The crucial difference of the adaptive procedure proposed in this paper with respect to AMR is that, in this paper, the model is coarsened or refined, and not the mesh.
\end{remark}

\begin{remark}
	For moment models in kinetic theory, it is shown that the magnitude of the last moment(s) can be used to detect non-equilibrium regions, which require more modeling complexity \cite{koellermeier_error_2019}, and near-equilibrium regions, where the model can be coarsened, similarly to the first criterion in \eqref{eq:residual_decrease_criteria} in Definition \ref{def:coarsening}. 
\end{remark}

\subsubsection{Example: \(\text{SWME}_{1}\) and \(\text{SWME}_{2}\)}
In this subsection, the sets of domain decomposition criteria for model-coarsening \eqref{eq:residual_decrease_criteria} and model-refining \eqref{eq:increase-criteria-physical} at \((t^*,x^*)\in[0,T]\times\Omega_x\) proposed in the previous subsection will be explicitly written for the case where the lower order is \(M_\mathrm{L}=1\) and the higher order is \(M_\mathrm{H}=2\). The \(\text{SWME}_{1}\) and the \(\text{SWME}_{2}\) are given by Equations \eqref{SWME-order1} and \eqref{SWME-order2}, respectively. In this case, the matrix \(\delta A_{1,2}(\vec{w}_2(t^*,x^*))\) and the vector \(\delta \vec{S}_{1,2}(\vec{w}_2(t^*,x^*))\) constructed in \ref{sec:appendix} are given by
\begin{equation}\label{eq:A-S-decomposition_example}
	\delta A_{1,2}(\vec{w}_2(t^*,x^*)) = 
	\begin{pmatrix}
		0 & 0 & 0 \\[5pt]
		-\frac{1}{5}\alpha_2^2 & 0 & 0 \\[5pt]
		-\frac{4}{5}\alpha_1\alpha_2 & 0 & \alpha_2
	\end{pmatrix}
	\quad \text{and} \quad 
	\delta \vec{S}_{1,2}(\vec{w}_2(t^*,x^*)) = 
	-\frac{\nu}{\lambda}
	\begin{pmatrix}
		0 \\
		\alpha_2 \\[5pt]
		3\alpha_2
	\end{pmatrix},
\end{equation}
respectively. Note that the dependence on \((t^*,x^*)\) is not explicitly written for readability. Clearly, \(\delta A_{1,2}(\vec{w}_2(t^*,x^*))\) and \(\delta \vec{S}_{1,2}(\vec{w}_2(t^*,x^*))\) vanish for vanishing moment \(\alpha_2\), as seen in \eqref{eq:SWME_diff_AMh_zeroLastMoment} and \eqref{eq:SWME_diff_SMh_zeroLastMoment}, respectively, and in \ref{sec:appendix}, for arbitrary orders \(M_\mathrm{H}\) and \(M_\mathrm{L}\). As mentioned before, this is a result of the hierarchical structure of the SWME \eqref{eq:SWME-compact}. In addition, \(\vec{S}_{1:2}(\vec{w}^{(0)}_{2}(t^*,x^*))\) is given by
\begin{equation}\label{eq:last-entry-source_example}
	\vec{S}_{1:2}(\vec{w}^{(0)}_{2}(t^*,x^*))=-5\frac{\nu}{\lambda}\left( u_m+\alpha_1 \right).
\end{equation}
Note that for \(\lambda \to 0\), the no-slip boundary condition at the bottom boundary effectively enforces a vanishing bottom velocity. Under the assumption that \(\alpha_2 = 0\), the bottom velocity in \(\mathrm{SWME}_2\) is given by \(u_m + \alpha_1\), which is precisely the quantity appearing on the right-hand side of Equation \eqref{eq:last-entry-source_example}. Consequently, Equation \eqref{eq:last-entry-source_example} remains consistent in the no-slip limit \(\lambda \to 0\). This strong enforcement of boundary conditions for the velocity profile was discussed explicitly in \cite{SWME}.

The domain decomposition criteria proposed in the previous subsection can now be written explicitly.
\paragraph{Model-coarsening criteria from \(M_\mathrm{H}=2\) to \(M_\mathrm{L}=1\)}
The residual \(\mathcal{R}_{:1}(t^*,x^*)\) \eqref{eq:res1} is given by
\begin{equation}\label{eq:domain-decomposition-criterion-1_reduced_example}
	\mathcal{R}_{:1}(t^*,x^*)=
	\left(
	0,\frac{2}{5}\alpha_2\partial_x(h\alpha_2),\frac{3}{5}\alpha_1\partial_x(h\alpha_2)
	\right)^T,
\end{equation}
and the residual \(\mathcal{R}_{1:2}(t^*,x^*)\) \eqref{eq:res2} reads
\begin{equation}\label{eq:domain-decomposition-criterion-2_example}
		\mathcal{R}_{1:2}(t^*,x^*)=
		\left(-\frac{2}{3}\alpha_1^2-2\alpha_2u_m-\frac{2}{7}\alpha_2^2\right)\partial_xh + 2\alpha_2\partial_x(hu_m) + \frac{1}{3}\alpha_1\partial_x(h\alpha_1) \left( u_m +\frac{3}{7}\alpha_2\right)\partial_x (h\alpha_2)
		+5\frac{\nu}{\lambda}\left(u_m+\alpha_1+\alpha_2+12\frac{\lambda}{h}\alpha_2\right).
\end{equation}
The model-coarsening criteria are thus given by
\begin{equation}\label{eq:decrease-criteria-physical_example}
	|h\alpha_2(t^*,x^*)|<\epsilon_w,\qquad
	\left\Vert
	\mathcal{R}_{:1}(t^*,x^*)
	\right\Vert_2 < \epsilon_{\mathcal{R}_{:1}},\qquad\Bigg\Vert
		\mathcal{R}_{1:2}(t^*,x^*)
		\Bigg\Vert_2
		< \epsilon_{\mathcal{R}_{1:2}},
\end{equation}
where \(\mathcal{R}_{:1}(t^*,x^*)\) and \(\mathcal{R}_{1:2}(t^*,x^*)\) are given by \eqref{eq:domain-decomposition-criterion-1_reduced_example} and \eqref{eq:domain-decomposition-criterion-2_example}, respectively. If all conditions in \eqref{eq:decrease-criteria-physical_example} are satisfied, then equations \eqref{eq:domain-decomposition-criterion-1_reduced} and \eqref{eq:domain-decomposition-criterion-2} are approximately fulfilled and the order is reduced at \((t^*,x^*)\) from higher order \(M_\mathrm{H}=2\) to lower order \(M_\mathrm{L}=1\).

\paragraph{Model-refinement criteria from \(M_\mathrm{L}=1\) to \(M_\mathrm{H}=2\)}
The model-refinement criteria \eqref{eq:increase-criteria-physical} are now written explicitly for the case \(M_\mathrm{L}=1\) and \(M_\mathrm{H}=2\):
\begin{equation}\label{eq:increase-criteria-physical_example}
\left|\vec{S}_{1:2}(\vec{w}^{(0)}_{2}(t^*,x^*))\right|>\kappa_S,\quad|\partial_x h(t^*,x^*)|>\kappa_{\partial}, \quad
|\partial_x \left(hu_m(t^*,x^*)\right)|>\kappa_{\partial}, \quad |\partial_x \left(h\alpha_1(t^*,x^*)\right)|>\kappa_{\partial},
\end{equation}
where \(\vec{S}_{1:2}(\vec{w}^{(0)}_{2}(t^*,x^*))\) is given by \eqref{eq:last-entry-source_example}. If one of the conditions in \eqref{eq:increase-criteria-physical_example} is satisfied, then the validity of Equation \eqref{eq:domain-decomposition-criterion-2} is not guaranteed and the order is increased from lower order \(M_\mathrm{L}=1\) to higher order \(M_\mathrm{H}=2\).

\subsection{Boundary interface coupling}\label{sec:spatial_coupling}
The second main challenge of the model-adaptive simulation of SWME models is the spatial coupling at the boundary interfaces, where two subdomains modeled by a different order SWME model border each other. In the context of finite volume schemes, the spatial coupling problem of two SWME models of a different order is the computation of the flux between two cells that have a different number of variables. In this paper, we propose two approaches that are extensions of the method proposed in \cite{PBC} to compute this boundary interface flux. The two approaches will be illustrated by considering one boundary interface and one canonical domain decomposition example. The resulting formulas can then be applied at each boundary interface for a general domain decomposition. 

\paragraph{Spatial discretization notation} 
Equation \eqref{eq:SWME-compact} is discretized on \(\Omega_x\) by means of a collocated, equidistant grid with grid cell size \(\Delta x\), yielding the semi-discretized cell-averaged moment vectors 
\begin{equation}\label{eq:semi-discrete_variables}
w_i=\left(h_i(t),(hu_m)_i(t),(h\alpha_1)_i(t),\ldots,(h\alpha_{M})_i(t)\right)^T \in \mathbb{R}^{M+2}
\end{equation} 
in the grid cells \(\mathcal{C}_i=[x_i-\Delta x/2,x_i+\Delta x/2]\) with equidistant cell centers \(x_i\), \(i=1,2,\ldots,N_x\), and discretized in time with time step \(\Delta t\), yielding the fully discretized cell-averaged moment vectors 
\begin{equation}\label{eq:discrete_variables}
w_i^n=\left(h_i^n,(hu_m)_i^n,(h\alpha_1)_i^n,\ldots,(h\alpha_{M})_i^n\right)^T \in \mathbb{R}^{M+2}
\end{equation} 
at discrete times \(t_n\), \(n=0,1,\ldots,N_t\). Note that the vector notation is dropped for simplicity and to distinguish the discretized moment vectors from the unknown functions described by Equation \eqref{eq:SWME-compact}. In particular, the subscript \(i\) in \(w_i\) denotes the cell-averaged value in the cell labeled \(i\) and should not be confused with the subscript \(M\) in \(\Vec{w}_M\) in Equation \eqref{eq:SWME-compact} that denotes the order of the moment model. 

\paragraph{Global finite volume discretization}
The overarching finite volume solver for the numerical simulation of Equation \eqref{eq:SWME-compact}, is a first-order time-splitting scheme. Equation \eqref{eq:SWME-compact} is split into a transport part and a source term part, yielding the two subproblems
\begin{alignat}{2}\label{eq:time-splitting1}
&\partial_t \vec{w}_M + A_M(\vec{w}_M)\partial_x\vec{w}_M \quad &=& \quad 0, \\[5pt]\label{eq:time-splitting2}
&\partial_t \vec{w}_M \quad &=& \quad S_M(\vec{w}_M).
\end{alignat} 
For the numerical simulation of the transport step \eqref{eq:time-splitting1}, we consider a finite volume scheme and choose the Polynomial Viscosity Method (PVM) \cite{castro_diaz_class_2012} notation, as common for applications of the SWME, because it is generally applicable to non-conservative systems and it includes well-known schemes such as the PRICE scheme \cite{PRICE}, which will be used for the numerical simulations in Section \ref{section:Numerics}. For the simulation of the transport step \eqref{eq:time-splitting1} of the \(\text{SWME}_M\) \eqref{eq:SWME-compact}, with a global order \(M\), the PVM reads
\begin{equation}\label{eq:PVM_fluctuationForm}
w_i^{n+1,*} = w_i^n - \frac{\Delta t}{\Delta x}\left( D_{i-\frac{1}{2}}^+ + D_{i+\frac{1}{2}}^- \right), \qquad i=1,\ldots,N_x,
\end{equation}
with fluctuations \(D_{i+\frac{1}{2}}^\pm=A_{\Phi}^\pm(w_i^n,w_{i+1}^n)\) given by
\begin{equation}\label{eq:fluct_definition}
A_{\Phi}^{\pm}(w_{\rm{l}},w_{\rm{r}}) = \frac{1}{2}\left( A_{\Phi}(w_{\rm{l}},w_{\rm{r}})\cdot (w_{\rm{r}}-w_{\rm{l}}) \pm Q_{\Phi}(w_{\rm{l}},w_{\rm{r}}) \cdot (w_{\rm{r}}-w_{\rm{l}}) \right),
\end{equation}
with generalized Roe linearization \(A_{\Phi}=A_{\Phi}(w_{\rm{l}},w_{\rm{r}})\) given by 
\begin{equation}\label{eq:generalizedRoe_NonLinear}
A_{\Phi}(w_{\rm{l}},w_{\rm{r}})\cdot (w_{\rm{r}}-w_{\rm{l}})=\int_{0}^1 A_M(\Phi(s;w_{\rm{l}},w_{\rm{r}}))\frac{\partial \Phi}{\partial s}(s;w_{\rm{l}},w_{\rm{r}})ds,
\end{equation}
and where the numerical viscosity matrix 
\begin{equation}\label{eq:numericalViscosity}
Q_{\Phi}(w_{\rm{l}},w_{\rm{r}})=P(A_{\Phi}(w_{\rm{l}},w_{\rm{r}}))
\end{equation}
is a function \(P(\cdot)\) of the generalized Roe linearization. Here, \(\Phi(s;w_{\rm{l}},w_{\rm{r}}):\mathbb{R}\mapsto\mathbb{R}^{M+2}\) denotes a path connecting the left state \(w_{\rm{l}}\in\mathbb{R}^{M+2}\) and the right state \(w_{\rm{r}}\in\mathbb{R}^{M+2}\) at the cell interface, such that \(\Phi(0;w_{\rm{l}},w_{\rm{r}})=w_{\rm{l}}\) and \(\Phi(1;w_{\rm{l}},w_{\rm{r}})=w_{\rm{r}}\). For future use, we state the linear path \(\Phi_{\text{lin}}(s;w_{\rm{l}},w_{\rm{r}})=(1-s)\cdot w_{\rm{l}} + s\cdot w_{\rm{r}} \).   
For the numerical simulations presented in Section \ref{section:Numerics}, the transport step \eqref{eq:time-splitting1} is numerically solved using the PRICE scheme \cite{PRICE}, which can be written as a PVM scheme \eqref{eq:PVM_fluctuationForm} with polynomial viscosity matrix
\begin{equation}\label{eq:viscosity-price}
Q_\Phi(w_{\rm{l}},w_{\rm{r}}) = \frac{\Delta x}{2\Delta t}I+\frac{\Delta t}{2\Delta x}A^2_\Phi(w_{\rm{l}},w_{\rm{r}}).
\end{equation} 
The source term step \eqref{eq:time-splitting2} is discretized using implicit Euler because of its unconditional stability. Because of the specific form of the source term in the SWME \eqref{eq:SWME-compact}, the implicit Euler method for \eqref{eq:time-splitting2} can be written as
\begin{equation}\label{eq:implicit-euler}
w^{n+1} = \left( I-\frac{\Delta t}{h^n}S \right)^{-1}w^{n+1,*},
\end{equation}
where \(w^{n+1,*}\) is the solution of the transport step \eqref{eq:PVM_fluctuationForm}, and where the matrix \(S\in\mathbb{R}^{(M+2)\times(M+2)}\) is defined in \cite{huang_equilibrium_2022} and can be computed offline to allow for an efficient integration of subproblem \eqref{eq:time-splitting2}. For the details, we refer to \cite{huang_equilibrium_2022}.

\begin{figure}[h!]
\centering
\includegraphics[width=0.8\linewidth]{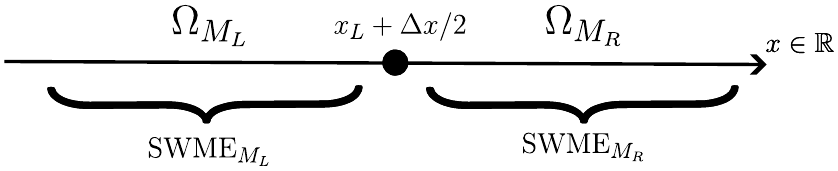}
\caption{Canonical domain decomposition example: the left lower-order subdomain \(\Omega_{M_L}\) is modeled by the lower-order \(\text{SWME}_{M_L}\) and the right higher-order subdomain \(\Omega_{M_R}\) is modeled by the higher-order \(\text{SWME}_{M_R}\), with \(M_L<M_R\). Modified from \cite{PBC}, Fig. 1.}
\label{fig:domain-decomposition}
\end{figure}

\paragraph{Canonical domain decomposition example} One particular domain decomposition situation is considered as a canonical example to describe the spatial coupling approaches discussed in the remainder of this section. The situation is illustrated in Figure \ref{fig:domain-decomposition}. Assume that a domain decomposition as outlined in Section \ref{sec:domain-decomposition} is performed at time \(t_n\) after obtaining the numerical solutions \(w_i^n\), for \(i=1,\ldots,N_x\), resulting in the updated model orders \(M_{L}^{n+1}\) and \(M_{R}^{n+1}\) in the subdomains \(\Omega_{M_L}\supset \mathcal{C}_L\) and \( \Omega_{M_R}\supset \mathcal{C}_R\), respectively, where \(L\in\{1,\ldots,N_x-1\}\) and \(R=L+1\) are cell indices. Note that in Section \ref{sec:domain-decomposition} the subscript '\(\mathrm{L}\)' in \(M_\mathrm{L}\) was used to denote the order of the lower-order model, while in this section the subscript '\(L\)' in the notation \(M_L^{n+1}\) is used to denote the order of the model used in the left subdomain \(\Omega_{M_L}\). The boundary interface is located at \(x=x_{L}+\Delta x/2\), between the cells \(\mathcal{C}_L\) and \(\mathcal{C}_R\), so cell \(\mathcal{C}_L\) is the cell left of the boundary interface \(x=x_{L}+\Delta x/2\) and cell \(\mathcal{C}_R\) is the cell right of the boundary interface \(x=x_{L}+\Delta x/2\). Moreover, assume, without loss of generality, that \(M_L^{n+1}<M_R^{n+1}\).

In the interior of the left subdomain \(\Omega_{M_L}\), the generalized Roe linearizations are given by 
\begin{equation}\label{eq:roeLinearizations-remainingInterfaces_left}
	A_{\Phi}(w_l,w_{l+1})\cdot (w_{l+1}-w_l)=\int_{0}^1 A_{M_L}(\Phi_{\text{lin}}(s;w_l,w_{l+1}))\frac{\partial \Phi_{\text{lin}}}{\partial s}(s;w_l,w_{l+1})ds, 
\end{equation}
whereas the generalized Roe linearizations in the interior of the right subdomain \(\Omega_{M_R}\) are given by
\begin{equation} \label{eq:roeLinearizations-remainingInterfaces_right}
	A_{\Phi}(w_r,w_{r+1})\cdot (w_{r+1}-w_r)=\int_{0}^1 A_{M_R}(\Phi_{\text{lin}}(s;w_r,w_{r+1}))\frac{\partial \Phi_{\text{lin}}}{\partial s}(s;w_r,w_{r+1})ds. 
\end{equation}
In the remainder of this section, generalized Roe linearizations at the boundary interface \(x=x_{L}+\Delta x/2\) are obtained. All generalized Roe linearizations are then inserted into the fluctuations \(D_{i+\frac{1}{2}}\) \eqref{eq:fluct_definition}, using \eqref{eq:numericalViscosity}, which fully determines the PVM scheme \eqref{eq:PVM_fluctuationForm} in the entire domain. 

\begin{remark}
In the remainder of this section, we will drop the superscripts \(n\) and \(n+1\) that indicate the discrete time for readability. However, note that the domain decomposition is time-dependent and that both the orders \(M_L\) and \(M_R\), and the boundary interface positions vary in time. 
\end{remark}
As illustrated in Figure \ref{fig:domain-decomposition}, cells \(\mathcal{C}_L\) and \(\mathcal{C}_R\) are neighboring cells, bordering the boundary interface at \(x=x_L+\Delta x/2\), that contain the left lower-order state variables 
\begin{equation}\label{eq:state-variables-CellI}
w_L=\left(h_L,(hu_m)_L,(h\alpha_1)_L,\ldots,(h\alpha_{M_L})_L\right)^T \in \mathbb{R}^{M_L+2}
\end{equation} 
and the right higher-order state variables 
\begin{equation}\label{eq:state-variables-CellI+1}
w_R=\left(h_R,(hu_m)_R,(h\alpha_1)_R,\ldots,(h\alpha_{M_R})_R\right)^T \in \mathbb{R}^{M_R+2},
\end{equation}
respectively. 
At the boundary interface, the PVM scheme \eqref{eq:PVM_fluctuationForm} needs to be adjusted to account for the fact that the left state \(w_L\in \mathbb{R}^{M_L+2}\) and the right state \(w_R\in \mathbb{R}^{M_R+2}\) are modeled by different transport matrices and contain a different number of variables. Two extensions of the classical PVM equations \eqref{eq:PVM_fluctuationForm} are proposed in the following two subsections: (1) The first approach uses an adaptation of the so-called Padded Buffer Cell (PBC) coupling, introduced for linear moment models of rarefied gases in \cite{PBC}. In the PBC coupling method, the lower-order state variable vector \(w_L\) is padded with additional moments, the so-called padded moments, such that it has the same length as the higher-order state variable vector \(w_R\). The padded moments are treated as additional variables. This approach allows for a smooth transition between the two different-order SWME models but can in general not be written in conservative form for conservative moment systems; (2) The second approach, called Path-Conservative Interface Flux (PCIF), is introduced in this paper. In the PCIF coupling method, the padded moments are set to zero, such that they can effectively be removed from the numerical scheme. The PCIF coupling method can be written in conservative form for conservative moment systems for the first \(M_L+2\) equations, but it results in a sharper transition between the two different-order SWME models, making the method less robust and more susceptible to small oscillations introduced by the coupling. Path-conservative formulations are important for nonconservative hyperbolic systems because weak solutions across discontinuities depend on the prescribed family of paths in phase space \cite{Maso1995DefinitionAW,Pares2006}, which may be particularly relevant for problems involving hydraulic jumps. Moreover, in the special case that the system contains a conservative subsystem, a path-conservative scheme is conservative for that subsystem \cite{CASTRO20088107}.

\subsubsection{Boundary interface flux: Padded Buffer Cell (PBC)}\label{section:PBC}
\begin{figure}
\centering
\includegraphics[width=\linewidth]{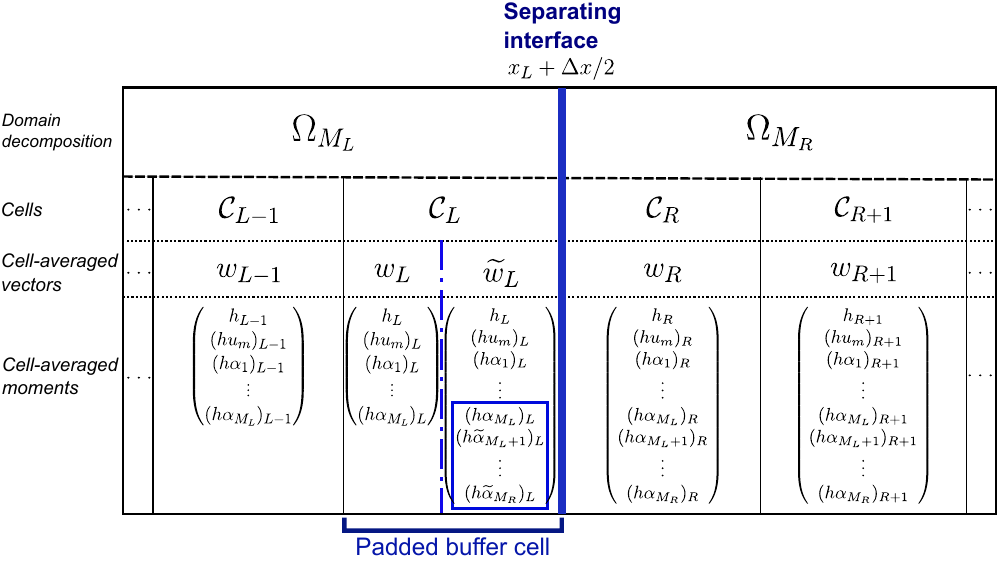}
\caption{Padded Buffer Cell with added moments \(\widetilde{\alpha}_{M_L\,+1},\ldots,\widetilde{\alpha}_{M_R}\) in cell \(\mathcal{C}_L\), yielding the padded vector \(\widetilde{w}_L\). The flux between cells \(\mathcal{C}_L\) and \(\mathcal{C}_R\) is computed using the padded vector \(\widetilde{w}_L\) instead of \(w_L\). Modified from \cite{PBC}, Fig. 2.}
\label{fig:domain-decomposition-discretized}
\end{figure}

In the PBC boundary interface coupling method \cite{PBC} applied to the non-linear and non-conservative SWME \eqref{eq:SWME-compact}, the lower-order \(\text{SWME}_{M_L}\) and the higher-order \(\text{SWME}_{M_R}\) are coupled by padding the left lower-order state variable vector \(w_L\in\mathbb{R}^{M_L+2}\) with additional moments  \(\left((\widetilde{\alpha}_{M_L\,+1})_L,\ldots,(\widetilde{\alpha}_{M_R})_L\right)^T\in \mathbb{R}^{M_R-M_L}\), yielding the so-called padded vector 
\begin{equation}\label{eq:padded_vector_SWME}
\widetilde{w}_L := \left(h_L, (hu_m)_L, (h\alpha_1)_L,\ldots, (h\alpha_{M_L})_L,(h\widetilde{\alpha}_{M_L\,+1})_L,\ldots,(h\widetilde{\alpha}_{M_R})_L\right)^T \in \mathbb{R}^{M_R+2},
\end{equation}
making it compatible with \(w_R\in\mathbb{R}^{M_R+2}\). This is illustrated in Figure \ref{fig:domain-decomposition-discretized}. 
The coupling is then defined by the generalized Roe linearization \eqref{eq:generalizedRoe_NonLinear} at the boundary interface between the cells \(\mathcal{C}_L\) and \(\mathcal{C}_R\), given by
\begin{equation}\label{eq:RoeLinearization_boundaryInterface_general}
A_{\Phi}(\widetilde{w}_L,w_R)\cdot (w_R-\widetilde{w}_L)=\int_{0}^1 A(\Phi(s;\widetilde{w}_L,w_R))\frac{\partial \Phi}{\partial s}(s;\widetilde{w}_L,w_R)ds,
\end{equation}
where the matrix \(A(\cdot)\in\mathbb{R}^{(M_{R}+2)\times(M_{R}+2)}\) and the path \(\Phi(\cdot)\in\mathbb{R}^{M_{R}+2}\) are yet to be defined. The PBC coupling uses the linear path \(\Phi(s;\widetilde{w}_L,w_R)=\Phi_\text{lin}(s;\widetilde{w}_L,w_R)\) and the higher-order system matrix \(A(\Phi(s;\widetilde{w}_L,w_R))=A_{M_R}(\Phi_\text{lin}(s;\widetilde{w}_L,w_R))\) along this path. The rightward fluctuation \(D_{L-1/2}^+\in\mathbb{R}^{M_{R}+2}\) \eqref{eq:fluct_definition} needs to implement a boundary condition for the padded moments \(\left((\widetilde{\alpha}_{M_L\,+1})_L,\ldots,(\widetilde{\alpha}_{M_R})_L\right)^T\), while the fluctuation \(D_{L-1/2}^-\in\mathbb{R}^{M_L+2}\) \eqref{eq:fluct_definition} uses \(w_L\) instead of the padded moment vector \(\widetilde{w}_L\), i.e.,
\begin{equation}\label{eq:flluctuation_wL-1_minusPlus}
D_{L-\frac{1}{2}}^-=A^-_{\Phi}(w_{L-1},w_L),\qquad
D_{L-\frac{1}{2}}^+=
\begin{pmatrix}
	A^+_{\Phi}(w_{L-1},w_L) \cr
	A_{\textrm{bound}}(w_{L-1},\widetilde{w}_L) \cr
\end{pmatrix}, 
\end{equation}
where \(A_{\textrm{bound}}(w_{L-1},\widetilde{w}_L)\in\mathbb{R}^{M_R-M_L}\) represents an interface boundary condition for the padded moments, which originates from the artificial boundary for the padded moments. We propose the boundary condition
\begin{equation}
A_{\textrm{bound}}(w_{L-1},\widetilde{w}_L)=a_{M_R-M_L}(\widetilde{w}_{L-1},\widetilde{w}_L)\cdot (\widetilde{w}_L-\widetilde{w}_{L-1}),
\end{equation}
where we defined \(\widetilde{w}_{L-1}\in\mathbb{R}^{M_R+2}\) as
\begin{equation}\label{eq:tilde-w_L-1}
\widetilde{w}_{L-1}:=\left(h_{L-1},\left( hu_m \right)_{L-1},\left(h\alpha_1\right)_{L-1},\ldots,\left(h\alpha_{M_L}\right)_{L-1},\left(h\widetilde{\alpha}_{M_L\,+1}\right)_L,\ldots,\left(h\widetilde{\alpha}_{M_R}\right)_L\right)^T,
\end{equation} 
and where \(a_{M_R-M_L}\in\mathbb{R}^{(M_R-M_L)\times(M_R+2)}\) are the last \(M_R-M_L\) rows of \(A^+_\Phi(\widetilde{w}_{L-1},\widetilde{w}_L)\) \eqref{eq:generalizedRoe_NonLinear}, computed using a linear path and using the system matrix of the \(\text{SWME}_{M_R}\). This specific boundary condition is chosen to avoid the introduction of waves at the boundary interface caused by the higher-order moments that are only defined in the right subdomain \(\Omega_{M_R}\). The PBC coupling is summarized as follows:
\begin{definition}[Padded Buffer Cell (PBC) coupling]\label{def:pbc}
Consider the canonical domain decomposition example illustrated in Figure \ref{fig:domain-decomposition}. Define the padded vector \(\widetilde{w}_L\) by \eqref{eq:padded_vector_SWME}. The Padded Buffer Cell (PBC) interface coupling for the computation of the updates for \(\widetilde{w}_L\) and \(w_R\) is given by the generalized Roe linearization
\begin{equation}\label{eq:RoeLinearization_boundaryInterface_PBC}
	A_{\Phi}(\widetilde{w}_L,w_R)\cdot (w_R-\widetilde{w}_L)=\int_{0}^1 A_{M_R}(\Phi_{\text{lin}}(s;\widetilde{w}_L,w_R))\frac{\partial \Phi_{\text{lin}}}{\partial s}(s;\widetilde{w}_L,w_R)ds    
\end{equation}
inserted in the fluctuations \(\mathcal{D}_{L+\frac{1}{2}}^{\pm}\) \eqref{eq:fluct_definition},
together with the fluctuations \(D_{L-\frac{1}{2}}^\pm\) defined in \eqref{eq:flluctuation_wL-1_minusPlus}. The generalized Roe linearizations at the remaining interfaces are given by \eqref{eq:roeLinearizations-remainingInterfaces_left} for \(l=1,\ldots,L-2\), and by \eqref{eq:roeLinearizations-remainingInterfaces_right}, for \(r=R,\ldots,N_x-1\).
\end{definition}
The PBC boundary interface coupling defined in Definition \ref{def:pbc} is a non-linear version of the existing PBC coupling introduced in \cite{PBC} for linear moment equations of rarefied gases. The PBC boundary interface coupling \(\eqref{eq:flluctuation_wL-1_minusPlus}\) can in general not be written in conservative form for a conservative system of moment equations, but it allows for a smooth transition between the two different order moment models.
\begin{remark}\label{remark:four_cells_comment}
For an efficient implementation of the PBC, we will require each subdomain in the domain decomposition to consist of at least four cells, such that the vectors \(\widetilde{w}_{L-1}\) and \(\widetilde{w}_L\) can be efficiently stored. Note that this smooths the domain decomposition. 
\end{remark}

\subsubsection{Boundary interface flux: Path-Conservative Interface Flux (PCIF)}
A general property of PVM schemes \eqref{eq:PVM_fluctuationForm} is that the schemes can be written in conservative form when applied to a conservative moment system, and thus it is desired that our adaptive finite volume scheme mimics this property. However, the PBC coupling proposed in the previous section based on \cite{PBC} does not satisfy this property. Note that the \(\text{SWME}_M\) \eqref{eq:SWME-compact} contains non-conservative terms for all \(M>0\), such that a treatment for these non-conservative terms is necessary. 
\begin{remark}
While we use the standard non-conservative SWME \cite{SWME} throughout this work, the general procedure proposed in this paper is also applicable to other hierarchical moment models. We are therefore using a general notation below. An example of a conservative system of moment equations is the Hermite Spectral Method (HSM)\cite{fan_accelerating_2020}, which is a linear system of PDEs describing the evolution of a rarefied gas. The HSM equations are simulated with a fixed domain decomposition and with the linear form of the PBC coupling \eqref{eq:flluctuation_wL-1_minusPlus}-\eqref{eq:RoeLinearization_boundaryInterface_PBC} in \cite{PBC}. The SWE \eqref{SWME-order0} can also be written in conservative form, but are included here in their non-conservative form. 
\end{remark}
To obtain an adaptive finite volume scheme that reduces to a conservative finite volume scheme for conservative moment equations, consider by means of example the general form of a conservative system of moment equations
\begin{equation}\label{eq:conservative-system}
\partial_t \vec{w}_M + \partial_x\vec{f}_M(\vec{w}_M) = \vec{0},
\end{equation}
describing the evolution of the conserved quantities \(\vec{w}_M\in\mathbb{R}^{M+2}\), for which the generalized Roe linearization \eqref{eq:generalizedRoe_NonLinear} is computed as 
\begin{equation}\label{eq:generalizedRoe_conservative}
A_{\Phi}(w_{\rm{l}}^n,w_{\rm{r}}^n)\cdot(w_{\rm{r}}^n-w_{\rm{l}}^n)=\vec{f}_M(w_{\rm{r}}^n)-\vec{f}_M(w_{\rm{l}}^n),
\end{equation}    
regardless of the path \(\Phi(s;w_\mathrm{l},w_\mathrm{r})\), so that the PVM scheme \eqref{eq:PVM_fluctuationForm} can be written as 
\begin{equation}\label{eq:FVM_conservative}
w_i^{n+1} = w_i^n-\frac{\Delta t}{\Delta x}\left(F_{i+\frac{1}{2}}^n-F_{i-\frac{1}{2}}^n\right),
\end{equation}
with numerical flux function based on a PVM method
\begin{equation}\label{eq:PVM_conservative_flux}
F_{i+\frac{1}{2}}^n=\frac{1}{2}\left(\vec{f}_M(w_i^n)+\vec{f}_M(w_{i+1}^n)-P(A_{\Phi}(w_i^n,w_{i+1}^n))\right).
\end{equation}
Note that we added the superscript \(n\) indicating the time discretization, but that this superscript will be dropped again for readability. We want to define the generalized Roe linearization at the boundary interface between the cells \(\mathcal{C}_L\) and \(\mathcal{C}_R\) such that the resulting finite volume scheme can be written in the conservative form \eqref{eq:FVM_conservative} for a conservative system of moment equations. We will therefore set the padded moments to zero: 
\begin{equation}\label{eq:cif_padded-moments}
\left((\widetilde{\alpha}_{M_L\,+1})_L,\ldots,(\widetilde{\alpha}_{M_R})_L\right)^T=\vec{0}\in\mathbb{R}^{M_R-M_L}.
\end{equation}
The domain decomposition notation introduced in Section \ref{sec:domain-decomposition} is applied here with \(M_\mathrm{L}=M_L\) and \(M_\mathrm{H}=M_R\). Then, we define a piecewise path \(\Phi_{\mathrm{pw}}(s;\widetilde{w}_L,w_R,s_0)\) resembling the Tuomi path \cite{toumi_weak_1992}:
\begin{equation}\label{eq:adaptive_path}
\Phi_{\mathrm{pw}}(s;\widetilde{w}_L,w_R,s_0)=\begin{cases}     \Phi_\mathrm{low}(s;\widetilde{w}_L,w_R)=\begin{pmatrix}
		\frac{s}{s_0}w_{R,:M_L}+\left(1-\frac{s}{s_0}\right)w_L \\
		(0,\ldots,0)^T
	\end{pmatrix}, & s\ \leq s_0, \\[15pt]
	\Phi_\mathrm{high}(s;\widetilde{w}_L,w_R)=\begin{pmatrix}
		w_{R,:M_L} \\
		\frac{s-s_0}{1-s_0}w_{R,M_L:M_R}
	\end{pmatrix}, & s > s_0,
\end{cases}
\end{equation}
where \(w_{R,:M_L}\in\mathbb{R}^{M_L+2}\) and \(w_{R,M_L:M_R}\in\mathbb{R}^{M_R-M_L}\) are defined as the first \(M_L+2\) entries of \(w_R\) and the last \(M_R-M_L\) entries of \(w_R\), respectively:
\begin{align}\label{eq:w_L+1,:Mi}
w_{R,:M_L} &:= \left(h_R,(hu_m)_R,(h\alpha_1)_R,\ldots,(h\alpha_{M_L})_R\right)^T\in\mathbb{R}^{M_L+2},\\[5pt]
w_{R,M_L:M_R} &:= \left((h\alpha_{M_L+1})_R,\ldots,(h\alpha_{M_R})_R\right)^T\in\mathbb{R}^{M_R-M_L}.
\end{align}
The parameter \(s_0\in[0,1]\) determines where the segment \(\Phi_\mathrm{low}(s;\cdot)\) ends and where the segment \(\Phi_\mathrm{high}(s;\cdot)\) starts. The choice of the piecewise path \(\Phi_{\mathrm{pw}}(s;\cdot)\) is motivated by the fact that the padded zeros are not computed variables. Therefore, it is intuitive to split the path in a segment that only considers the first \(M_L+2\) variables and a segment that only considers the last \(M_R-M_L\) variables while keeping the first \(M_L+2\) variables fixed. Note that the path \eqref{eq:adaptive_path} is continuous but not differentiable in \(s=s_0\). The integral in the generalized Roe linearization \eqref{eq:RoeLinearization_boundaryInterface_general} is effectively split into an integral along the segment \(\Phi_\mathrm{low}(s;\cdot)\) and an integral along the segment \(\Phi_\mathrm{high}(s;\cdot)\), and the higher-order system matrix \(A_{M_R}(\cdot)\) is evaluated along the path in both segments \(\Phi_\mathrm{low}(s;\cdot)\) and \(\Phi_\mathrm{high}(s;\cdot)\). The PCIF boundary coupling method is then defined in the following definition:
\begin{definition}[Path-Conservative Interface Flux (PCIF) coupling]\label{def:PCIF}
Consider the canonical domain decomposition example illustrated in Figure \ref{fig:domain-decomposition}. Define the padded vector \(\widetilde{w}_L\) by \eqref{eq:padded_vector_SWME}, with padded moments \(\left((\widetilde{\alpha}_{M_L\,+1})_L,\ldots,(\widetilde{\alpha}_{M_R})_L\right)^T=\vec{0}\) equal to zero according to \eqref{eq:cif_padded-moments}. The Path-Conservative Interface Flux (PCIF) interface coupling for the computation of the updates for \(\widetilde{w}_L\) and \(w_R\) is given by the generalized Roe linearization
\begin{align}\label{eq:GeneralizedRoe_split_inserted}
	\begin{split}
		&A_{\Phi}(\widetilde{w}_L,w_R)\cdot (w_R-\widetilde{w}_L)=    \frac{1}{1-s_0}\left[\int_{s_0}^{1}
		\begin{pmatrix}
			A_{:M_L,M_L:M_R}(\Phi_\mathrm{high}(s;\widetilde{w}_L,w_R)) \\[5pt]
			A_{M_L:M_R,M_L:M_R}(\Phi_\mathrm{high}(s;\widetilde{w}_L,w_R))
		\end{pmatrix}ds\right]\cdot w_{R,M_L:M_R} \\[5pt]
		& \qquad\qquad +\frac{1}{s_0}\left[\int_{0}^{s_0} 
		\begin{pmatrix}
			A_{M_L}\left(\frac{s}{s_0}w_{R,:M_L}+\left(1-\frac{s}{s_0}\right)w_L\right) \\[5pt]
			A_{M_L:M_R,:M_L}\left(\frac{s}{s_0}w_{R,:M_L}+\left(1-\frac{s}{s_0}\right)w_L\right)
		\end{pmatrix}ds\right]\cdot\left(w_{R,:M_L}-w_L\right),
	\end{split}
\end{align}
inserted in the fluctuations \(\mathcal{D}_{L+\frac{1}{2}}^{\pm}\) \eqref{eq:fluct_definition} and with the last \(M_R-M_L\) entries of \(\mathcal{D}_{L+\frac{1}{2}}^-\) set to zero. The generalized Roe linearizations at the remaining interfaces are given by \eqref{eq:roeLinearizations-remainingInterfaces_left} for \(l=1,\ldots,L-1\), and by \eqref{eq:roeLinearizations-remainingInterfaces_right} for \(r=R,\ldots,N_x-1\).
\end{definition}
Note that, to fix the padded moments \(\left((\widetilde{\alpha}_{M_L\,+1})_L,\ldots,(\widetilde{\alpha}_{M_R})_L\right)^T=\vec{0}\), the last \(M_R-M_L\) entries of \(\mathcal{D}_{L+\frac{1}{2}}^-\) are set to zero. By numerically solving the integrals in \eqref{eq:GeneralizedRoe_split_inserted} with a separate quadrature rule and then adding the numerical approximations, the quadrature points are distributed independently along the two segments \(\Phi_\mathrm{low}(s;\cdot)\) and \(\Phi_\mathrm{high}(s,\cdot)\). The parameter \(s_0\) can then be removed from \eqref{eq:GeneralizedRoe_split_inserted} by transforming both integration intervals \([0,s_0]\) and \([s_0,1]\) to the interval \([0,1]\), such that the value of \(s_0\) does not influence the PVM and does not need to be specified.
\begin{proposition}[PCIF coupling applied to conservative system]\label{conservative-system}
Consider the canonical domain decomposition example illustrated in Figure \ref{fig:domain-decomposition}. Consider the conservative system of moment equations \eqref{eq:conservative-system} and a corresponding finite volume scheme of the type \eqref{eq:FVM_conservative} with a numerical flux function \eqref{eq:PVM_conservative_flux} that is well defined between two cells modeled by a SWME model of the same order. The PCIF boundary interface coupling method can be written in the form \eqref{eq:FVM_conservative} for the system of equations \eqref{eq:conservative-system} for the first \(M_L+2\) variables by only modifying the flux \(F_{L+\frac{1}{2}}^n\).
\end{proposition}
\begin{proof}
Writing \eqref{eq:FVM_conservative} for the cells bordering on the interface at \(x_{L+1/2}\) for the first \(M_L+2\) variables, we obtain
\begin{align}\label{eq:FVM-conservative-atInterfac1}
	w_L^{n+1} &= w_L^n-\frac{\Delta t}{\Delta x} \left( \widetilde{F}_{L+\frac{1}{2}}^n - F_{L-\frac{1}{2}}^n \right), \\\label{eq:FVM-conservative-atInterfac2}
	w_{R,:M_L}^{n+1} &= w_{R,:M_L}^n-\frac{\Delta t}{\Delta x} \left( F_{I+\frac{3}{2}}^n - \widetilde{F}_{L+\frac{1}{2}}^n \right).
\end{align}
Note that we are only modifying the flux \(\widetilde{F}_{L+\frac{1}{2}}^n\) across the interface at \(x_{L+\frac{1}{2}}\) and that \(F_{L-\frac{1}{2}}^n\) and \(F_{I+\frac{3}{2}}^n\) are given by the standard numerical flux \eqref{eq:PVM_conservative_flux}. The discrete time superscripts are removed again in the following. Analogously to the notation introduced in Section \ref{sec:domain-decomposition}, the flux \(\vec{f}_{M_R}(w_R)\) of the system of order \(M_R\) is decomposed as
\begin{equation}
	\vec{f}_{M_R}(w_R) = 
	\begin{pmatrix}
		\vec{f}_{:M_L} (w_R) \\[5pt]
		\vec{f}_{M_L:M_R}(w_R)
	\end{pmatrix}, \quad \vec{f}_{:M_L} (w_R)\in\mathbb{R}^{M_L+2},\quad\vec{f}_{M_L:M_R} (w_R)\in\mathbb{R}^{M_R-M_L}.
\end{equation}
To compute the generalized Roe linearization \eqref{eq:GeneralizedRoe_split_inserted} for the conservative system \eqref{eq:conservative-system}, we have to evaluate \(\vec{f}_{M_R}(\cdot)\) in \(\Phi_{\mathrm{pw}}(0;\widetilde{w}_L,w_R)\) and \(\Phi_{\mathrm{pw}}(1;\widetilde{w}_L,w_R)\). Since
\begin{equation}\label{eq:adaptivePath_evaluationEndPoints}
	\Phi_{\mathrm{pw}}(0;\widetilde{w}_L,w_R)=
	(w_L,
	\vec{0})^T, \qquad \Phi_{\mathrm{pw}}(1;\widetilde{w}_L,w_R)=w_R,
\end{equation}
we obtain 
\begin{equation}
	f(\Phi_{\mathrm{pw}}(0;\widetilde{w}_L,w_R))=
	\begin{pmatrix}
		\vec{f}_{M_L}(w_L)\\[5pt]\vec{f}_{M_L:M_R}(\widetilde{w}_L)
	\end{pmatrix},\quad f(\Phi_{\mathrm{pw}}(1;\widetilde{w}_L,w_R))=\vec{f}_{M_R}(w_R).
\end{equation} 
The generalized Roe linearization at the boundary interface can then be computed as
\begin{equation}\label{eq:generalizedRoe_conservative_AtInterface}
	A_{\Phi}(\widetilde{w}_L,w_R)\cdot(w_R-\widetilde{w}_L)=
	\vec{f}_{M_R}(w_R)-
	\begin{pmatrix}
		\vec{f}_{M_L}(w_L)\\[5pt]\vec{f}_{M_L:M_R}(\widetilde{w}_L)
	\end{pmatrix},
\end{equation}  
with corresponding numerical viscosity matrix \(P(A_{\Phi}(\widetilde{w}_L,w_R))\). The generalized Roe linearization \eqref{eq:generalizedRoe_conservative_AtInterface} is only used in the fluctuation \(\mathcal{D}_{L+\frac{1}{2}}^+\in\mathbb{R}^{M_R+2}\), which is defined by inserting the generalized Roe linearization \eqref{eq:generalizedRoe_conservative_AtInterface} in the general formula \eqref{eq:fluct_definition}. The fluctuation \(\mathcal{D}_{L+\frac{1}{2}}^-\in\mathbb{R}^{M_L+2}\) is computed using the standard generalized Roe linearization
\begin{equation}\label{eq:generalizedRoe_conservative_AtInterface_negative}
	\widetilde{A}_{\Phi}(\widetilde{w}_L,w_R)\cdot(w_R-\widetilde{w}_L)=
	\vec{f}_{:M_L}(w_R)-
	\vec{f}_{M_L}(w_L)
\end{equation}  
in the fluctuation
\begin{equation}
	A_{\Phi}^{-}(\widetilde{w}_L,w_R) = \frac{1}{2}\left( \widetilde{A}_{\Phi}(\widetilde{w}_L,w_R)\cdot (w_R-\widetilde{w}_L) - \widetilde{Q}_{\Phi}(\widetilde{w}_L,w_R) \cdot (w_R-\widetilde{w}_L) \right),
\end{equation}
with numerical viscosity matrix \(\widetilde{Q}_{\Phi}(\widetilde{w}_L,w_R) =P(\widetilde{A}_{\Phi}(\widetilde{w}_L,w_R) )\). Note that the generalized Roe linearization is the same in the fluctuations \(\mathcal{D}_{L+\frac{1}{2}}^-\) and \(\mathcal{D}_{L+\frac{1}{2},:M_L}^+\in\mathbb{R}^{M_L+2}\) (the first \(M_L+2\) entries of \(\mathcal{D}_{L+\frac{1}{2}}^+\)). It follows that the scheme can be written in conservative form \eqref{eq:FVM_conservative} for the first \(M_L+2\) variables, with boundary interface flux function 
\begin{equation}\label{eq:fluxFunction_adaptive}
	\widetilde{F}^n_{L+\frac{1}{2},:M_L} = \frac{1}{2}\left(\vec{f}_{M_L}(w_L^n)+\vec{f}_{:M_L}(w_R^n) - P(A_{\Phi}(\widetilde{w}_L^n,w_R^n))\right)    
\end{equation}
for the first \(M_L+2\) variables.
\end{proof}
\begin{remark}
From the above arguments, it follows that if the padded moments of the padded vector \(\widetilde{w}_L\) are given by \eqref{eq:cif_padded-moments}, then \eqref{eq:RoeLinearization_boundaryInterface_general} reduces to conservative form \eqref{eq:FVM_conservative} for the conservative system \eqref{eq:conservative-system} regardless of the path \(\Phi(s;\widetilde{w}_L,w_R)\).
\end{remark}

The difference between the PBC coupling discussed in Section \ref{section:PBC} and the PCIF coupling introduced in this section is that in the PBC coupling the evolution of the padded values is computed with the PVM equations, while in the PCIF coupling the padded values are identically zero and can be effectively removed from the equations. In the numerical simulations of the PBC coupling presented in Section \ref{section:Numerics}, the padded buffer cells were initialized by setting them equal to their neighbour:
\begin{equation}\label{eq:padded_vector_SWME_PBC}
	\widetilde{w}_L := \left(h_L, (hu_m)_L, (h\alpha_1)_L,\ldots, (h\alpha_{M_L})_L,(h\alpha_{M_L\,+1})_R,\ldots,(h\alpha_{M_R})_R\right)^T \in \mathbb{R}^{M_R+2},
\end{equation}
so that the artificial jumps in the higher-order moments do not contribute to the variable updates.

\section{Numerical simulations}\label{section:Numerics}
In this section, the two spatial coupling methods PBC \eqref{eq:flluctuation_wL-1_minusPlus}-\eqref{eq:RoeLinearization_boundaryInterface_PBC} and PCIF \eqref{eq:GeneralizedRoe_split_inserted} proposed in Section \ref{sec:spatial_coupling} are numerically tested, using the domain decomposition criteria proposed in Section \ref{sec:domain-decomposition}. First, we will consider the collision of a dam break wave with a smooth wave, see Figure \ref{fig:initial_height} for the initial water height, for different friction parameter combinations. This is a combination of commonly considered test cases for the SWME, see for example \cite{HSWME,SWME}. At the dam break wave and the smooth wave, we expect the domain decomposition to indicate the need for a higher-order SWME model, while in the regions away from the dam break wave and the smooth wave, we expect the domain decomposition criteria to indicate that a lower-order SWME model can be used. We further expect that the eventual collision of the dam break wave and the smooth wave increases the flow complexity and thus requires time adaptivity, such that a higher-order SWME model is needed for accurate simulations. The accuracy and runtime of the two spatial coupling methods are compared with the accuracy and the runtime of an accurate but slow higher-order SWME model and a fast but inaccurate lower-order SWME model, as well as with a reference solution.

Finally, an additional test case involving a classical dam-break configuration is considered. For this test case, the errors and the runtimes of the A-SWME-PBC and the A-SWME-PCIF are compared with a set of fixed-order SWME models. Analogous to the first test case, we expect the domain decomposition to increase the order in regions reached by the dam-break waves.

The numerical simulations are based on a dimensionless form of the SWME \eqref{eq:SWME-compact}, analogous to \cite{SWME}. Equation \eqref{eq:SWME-compact} is nondimensionalized using a particular scaling, after which the resulting system is solved using dimensionless friction coefficients. To keep the notation light, we use the same notation for both the dimensional and dimensionless systems. For details, we refer to \cite{SWME}.

\paragraph{Notation} In the remainder of this paper, we will denote the adaptive model with PBC coupling by the abbreviation A-SWME-PBC and the adaptive model with PCIF coupling by the abbreviation A-SWME-PCIF.

\subsection{Numerical evaluation of domain decomposition criteria}\label{section:numerical-evaluation-criteria}
The model-coarsening criteria \eqref{eq:residual_decrease_criteria} and the model-refinement criteria \eqref{eq:increase-criteria-physical} need to be evaluated on the grid in each cell \(\mathcal{C}_i\), \(i=1,\ldots,N_x\), for each time step. This is computationally cheap compared to the full finite volume update \eqref{eq:time-splitting1}-\eqref{eq:time-splitting2}. For a practical implementation, only an increase or a decrease in order of one is applied. This corresponds to \(M_\mathrm{H}=M+1\) and \(M_\mathrm{L}=M\), for some \(M\in\mathbb{N}\), in the notation introduced in Section \ref{sec:domain-decomposition}. The efficient implementation of a more general decrease or increase in order is left to future work. 

For the numerical evaluation of the model-coarsening criteria \eqref{eq:residual_decrease_criteria} in a cell \(\mathcal{C}_i\) that is currently modeled by the \(\text{SWME}_{M_i+1}\), assume that the numerical solution \(w_i^n\) has been obtained up to discrete time \(t_n\) and that the current time step \(t_n\mapsto t_{n+1}\) aims at computing the numerical solution \(w_i^{n+1}\) at discrete time \(t_{n+1}\). The residual \(\mathcal{R}_{:M_i}(t_n,x_i)\) \eqref{eq:res1} is numerically approximated using the last column of the fluctuation \(A_{\Phi}^+(w_{i-1}^n,w_{i}^n)\), denoted by \(A_{\Phi,M_i+1}^+(w_{i-1}^n,w_{i}^n)\), and the last column of the fluctuation \(A_{\Phi}^-(w_i^n,w_{i+1}^n)\), denoted by \(A_{\Phi,M_i+1}^-(w_{i}^n,w_{i+1}^n)\), computed in the previous time step, divided by the grid cell size \(\Delta x\):
\begin{equation}\label{eq:res1_approximated}
	\mathcal{R}_{:M_i}(t_n,x_i)\approx \frac{A_{\Phi,M_i+1}^+(w_{i-1}^n,w_{i}^n)+ A_{\Phi,M_i+1}^-(w_{i}^n,w_{i+1}^n)}{\Delta x}.
\end{equation}
Note that \(\mathcal{R}_{M_i:M_i+1}(t_n,x_i)\) \eqref{eq:res2} describes the time evolution of the highest-order variable \((h\alpha_{M_i+1})_i^n\) at time \(t_n\). Instead of approximating the spatial fluctuations and the source term in \eqref{eq:res2}, we will approximate the time evolution directly by approximating the time derivative of the highest-order variable \((h\alpha_{M_i+1})_i^n\) at time \(t_n\) using a finite difference approximation:
\begin{equation}\label{eq:res2_approximated}
	\mathcal{R}_{M_i:M_i+1}(t_n,x_i)\approx \frac{(h\alpha_{M_i+1})_i^n-(h\alpha_{M_i+1})_i^{n-1}}{\Delta t}.
\end{equation}
Note that special care needs to be taken when computing \eqref{eq:res1_approximated} and \eqref{eq:res2_approximated} to treat missing values as a result of the time- or spatially-adaptive subdomains. The approximations \eqref{eq:res1_approximated} and \eqref{eq:res2_approximated} are then inserted in the model-coarsening criteria \eqref{eq:residual_decrease_criteria}.
For the numerical evaluation of the model-refinement criteria \eqref{eq:increase-criteria-physical} in a cell \(\mathcal{C}_i\) that is currently modeled by the \(\text{SWME}_{M_i}\), the model-refinement criteria in \eqref{eq:increase-criteria-physical} are approximated on the grid:
\begin{equation}\label{eq:increase-criteria-physical_discrete}
	|S_{M_i:M_i+1}(w_{i}^{n,(0)})|>\kappa_S,\quad |\Delta h_i^n|>\kappa_{\partial},\quad |\Delta (hu_m)_i^n|>\kappa_{\partial}, \quad|\Delta (h \alpha_1)_i^n|>\kappa_{\partial},\quad\cdots,\quad |\Delta (h\alpha_{M_i})_i^n|>\kappa_{\partial},
\end{equation}
with \(w_{i}^{n,(0)}:=\left((w_i^n)^T,0\right)^T\), and where the partial derivatives in \eqref{eq:increase-criteria-physical} have been replaced by central difference approximations denoted by \(\Delta h_i^n\), \(\Delta \left(hu_m\right)_i^n\) and \(\Delta \left(h\alpha_j\right)_i^n\), \(j=1,\ldots,M_i\). The specific values for the tolerances were chosen based on heuristic testing and are summarized in Table \ref{tab:tolerances}. 
\begin{table}[H]
	\begin{center}
		\begin{tabular}{ | c || c || c || c || c | } 
			\hline
			\(\epsilon_w =0.001\) &  \(\epsilon_{\mathcal{R}_{:M_i}} =0.005\)  &  \(\epsilon_{\mathcal{R}_{M_i:M_i+1}} =0.25\)  &  \(\kappa_S =0.5\)  &  \(\kappa_{\partial} =0.2\) \\[3pt]
			\hline
		\end{tabular}
		\caption{Chosen tolerances in \eqref{eq:residual_decrease_criteria} and \eqref{eq:increase-criteria-physical_discrete} for the adaptive simulation.}
		\label{tab:tolerances}
	\end{center}
\end{table}
The extension to higher-order methods (in time and space) requires an adaptation of the error estimation for the spatial terms as well as for the time integration over the time step. This is left for future work.

The algorithm for the adaptive simulation is summarized in Algorithm \ref{alg:adaptive-simulation-swme}. First, the order in each cell is initialized and set to a maximum order \(M_{max}\). For the simulation results shown in this section, the maximum order was chosen to be \(M_{max}=5\). At each discrete time \(t_n\) and in each grid cell \(\mathcal{C}_i\), the transport and friction updates are computed according to equations \eqref{eq:time-splitting1} and \eqref{eq:time-splitting2}, respectively. Then, the model-refinement criteria \(\eqref{eq:increase-criteria-physical_discrete}\) are evaluated to identify subdomains in which the order of the SWME model should be increased. Then, in cells that have not been identified for model-refinement, the model-coarsening criteria \eqref{eq:residual_decrease_criteria} are numerically evaluated to identify subdomains in which the order of the SWME model can be reduced.

\begin{algorithm}
	\caption{Algorithm for the model-adaptive simulation of SWME}\label{alg:adaptive-simulation-swme}
	\begin{algorithmic}
		\Require $M_{max}\geq 0$
		\For{$i \gets 1$ to $N_x$}                    
		\State {$M[i]$ $\gets$ {$M_{max}$}}\Comment{Initialize cell-wise orders} 
		\EndFor
		\State $t \gets 0$
		\While{$t \leq t_{end}$}
		\For{$i \gets 1$ to $N_x$}           
		\State Compute time step size \(\Delta t\)
		\State Solve transport step \eqref{eq:time-splitting1} with time step \(\Delta t\)
		\State Solve friction step \eqref{eq:time-splitting2} with time step \(\Delta t\)
		\If{\eqref{eq:increase-criteria-physical_discrete} is satisfied and $M[i]<M_{max}$}
		\State $M[i]$ $\gets$ $M[i]+1$  
		\ElsIf{\eqref{eq:residual_decrease_criteria} is numerically satisfied and $M[i]>0$}\Comment{Using \eqref{eq:res1_approximated}-\eqref{eq:res2_approximated}}
		\State $M[i]$ $\gets$ $M[i]-1$  
		\EndIf
		\EndFor
		\State $t \gets t+\Delta t$
		\EndWhile
	\end{algorithmic}
\end{algorithm}

\subsection{Reference solution}\label{section:referenceSolution}
Reference solutions are obtained using the software framework employed in \cite{SWME} to solve the reference system \eqref{eq:ref1}-\eqref{eq:ref2} numerically. Equations \eqref{eq:ref1}-\eqref{eq:ref2} are split into a (hyperbolic) transport part and a diffusive part. The transport part is discretized in space with a third-order finite volume scheme and advanced in time using a third-order explicit Runge-Kutta method. The diffusive part is discretized in space using finite differences and integrated in time with a third-order implicit Runge-Kutta method. We refer to \cite{SWME} for details.

\subsection{Numerical results: dam-break wave colliding with a smooth wave}
\begin{figure}[htb!]
	\centering
	\includegraphics[width=0.5\linewidth]{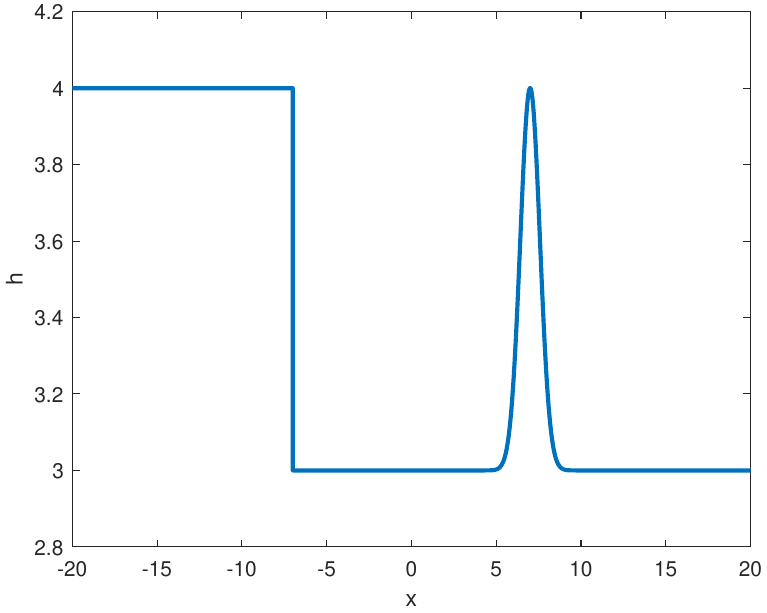}
	\caption{Initial height profile of the test case displayed in Table \ref{tab:setup-general}.}
	\label{fig:initial_height}
\end{figure}

In this subsection, the results of the numerical simulation of a dam-break wave colliding with a smooth wave for three friction cases are shown. The setup of the test case is displayed in Table \ref{tab:setup-general}. The initial height \(h(0,x)\) is a combination of a dam and a smooth wave, see Figure \ref{fig:initial_height}. The initial velocity \(u(0,x,\zeta)\) is linear in the vertical variable and does not vary in the horizontal variable. The time step size \(\Delta t\) at each time step \(n\) is computed by the CFL-condition, based on the system matrix of the highest order that was observed in the spatial cells, and was taken to be constant in all grid cells. The reference simulations have been obtained using a grid with a grid cell size of \(\Delta x_{\text{ref}}=0.02\) in \(x-\)direction and \(\Delta \zeta_{\text{ref}} = 0.005\) in \(\zeta -\)direction, with a CFL number of 0.5.  
\begin{table}[htb!]
	\begin{center}
		\begin{tabular}{ | m{10em} || m{6.5cm} | } 
			\hline
			Friction parameters & \((\nu,\lambda)\in \{ (0.1,0.1), (1,0.1),(0.1,1) \}\) \\[4pt] \hline 
			Spatial domain & \(x \in [-20,20]\)\\[4pt]
			\hline
			grid cell size & \(\Delta x = 0.004\)\\[4pt]
			\hline
			End time & \(t=5\) \\[4pt]
			\hline
			Initial condition & \(
			h(0,x) = 
			\begin{cases}
				4, &  x \leq -7 \\
				3+e^{-1.5(x-7)^2}, &  x > -7
			\end{cases}
			\) \\[5pt] 
			& \(u_m(0,x)=0.05\) \\[5pt]
			& \(\alpha_1=-0.01\) \\[5pt]
			& \(\alpha_i=0, \quad i>1\) \\[5pt]
			\hline
			Initial velocity profile & \(u(0,x,\zeta) = 0.04+0.02\zeta \) \\[4pt]
			\hline
			CFL number & 0.5 \\[4pt]
			\hline
		\end{tabular}
		\caption{Numerical setup for collision of a dam break wave with a smooth wave.}
		\label{tab:setup-general}
	\end{center}
\end{table}
For each friction case, four models are simulated: (1) The lower-order model (either the \(\text{SWME}_1\) or the \(\text{SWME}_2\), depending on the friction case); (2) the \(\text{SWME}_5\) model, further referred to as the higher-order model; (3) the A-SWME-PBC; (4) the A-SWME-PCIF. The friction parameters \(\lambda\) (slip length) and \(\nu\) (kinematic viscosity) play an important role in the shape of the vertical velocity, because their magnitudes are closely related to the magnitude of the source terms that drive the vertical variability in the velocity profile. We therefore expect to see qualitative differences in the performances of the adaptive schemes for the different friction cases defined in Table \ref{tab:setup-general}. The accuracy and the runtime of the four models will be compared, as well as the domain decompositions using the two adaptive schemes. The higher-order \(\text{SWME}_5\) showed sufficient convergence towards the reference solution, such that the reference solution is not shown in the plots and only used to compute relative model errors.

\begin{figure}[htb!]
	\centering
	\begin{subfigure}[b]{0.475\textwidth}
		\centering
		\includegraphics[width=\textwidth]{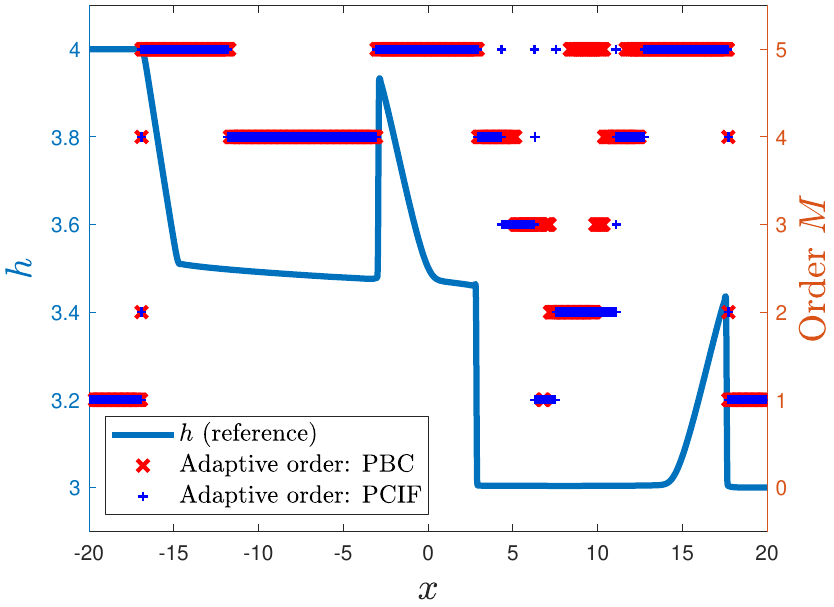}
		\caption{Water height \(h\) alongside adaptive orders of PBC and PCIF}
		\label{fig:lambda1.0_nu0.1_t5_h_full}
	\end{subfigure}
	\hfill
	\begin{subfigure}[b]{0.45\textwidth}
		\centering
		\includegraphics[width=\textwidth]{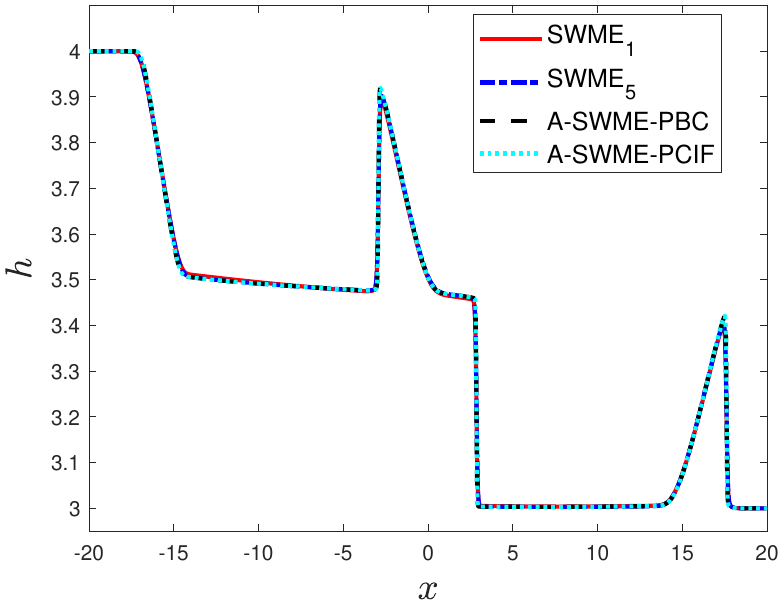}
		\caption{Water height \(h\)}
		\label{fig:lambda1.0_nu0.1_t5_h}
	\end{subfigure}
	\hfill
	\begin{subfigure}[b]{0.45\textwidth}
		\centering
		\includegraphics[width=\textwidth]{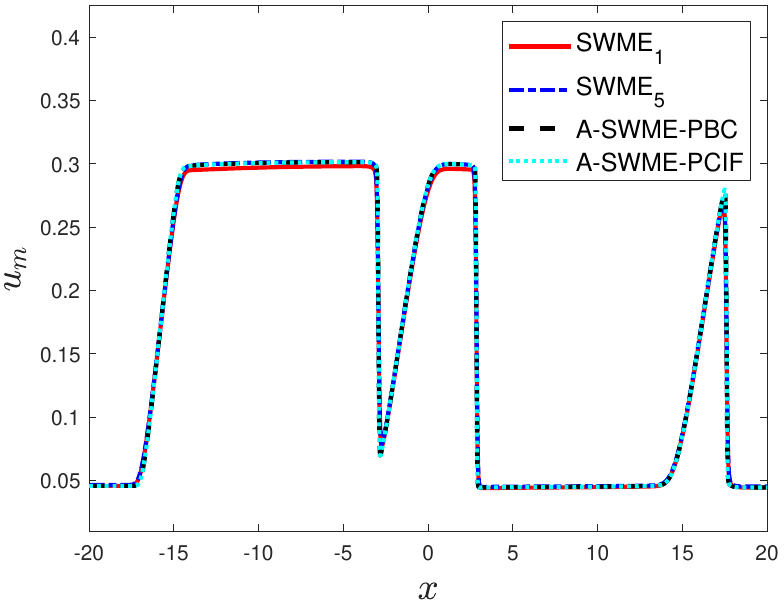}
		\caption{Mean velocity \(u_m\)}
		\label{fig:lambda1.0_nu0.1_t5_u_full}
	\end{subfigure}
	\hfill
	\begin{subfigure}[b]{0.45\textwidth}
		\centering
		\includegraphics[width=\textwidth]{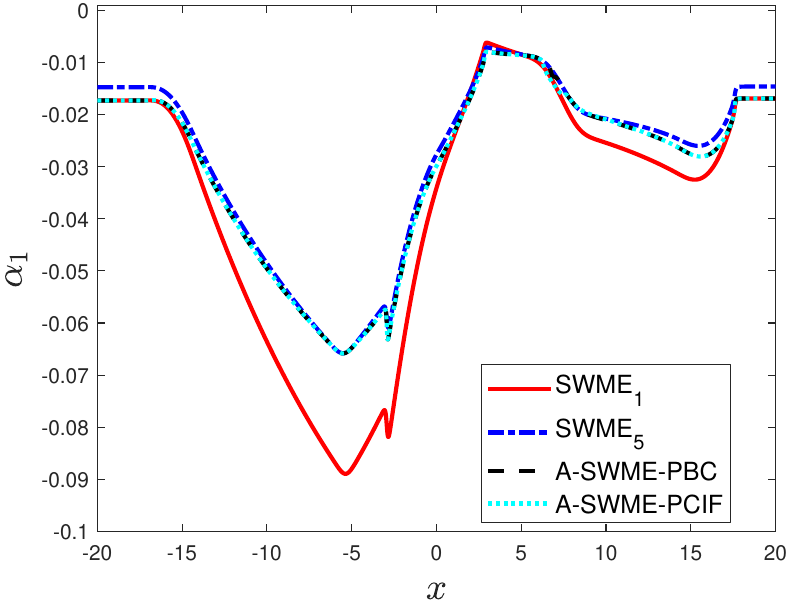}
		\caption{First moment \(\alpha_1\)}
		\label{fig:lambda1.0_nu0.1_t5_alpha1_full}
	\end{subfigure}
	\caption{Collision of a dam break wave with a smooth wave with friction parameters \(\lambda=1\) and \(\nu=0.1\) (friction case 1) at time \(t=5\). (a): Solid blue line is the simulated water height from the reference solution. The red crosses and the blue plus signs are the orders in each cell using the PBC coupling and the PCIF coupling, respectively. (b)-(d): The lower-order model is the \(\text{SWME}_1\) (solid red line). The higher-order model is the \(\text{SWME}_5\) model (blue dash-dotted line). The A-SWME-PBC (black dashed line) and the A-SWME-PCIF (cyan dotted line) yield similar results, close to the higher-order model, for the height \(h\), mean velocity \(u_m\), and first moment \(\alpha_1\).}
	\label{fig:lambda1.0_nu0.1_t5}
\end{figure}

\paragraph{Friction case 1: \(\lambda = 1\) and \(\nu = 0.1\)} 
In this friction case, the slip length \(\lambda\) is large compared to the viscosity \(\nu\). The large slip indicates that the velocity profile is not relaxing toward zero so quickly at the bottom, whereas the small viscosity is unlikely to result in a strongly variable vertical velocity profile because of little internal friction. The adaptive models resulted in a minimum order \(M=1\), so the \(\text{SWME}_1\) model was chosen as the lower-order model for fair comparison. The results at time \(t=5\) are shown in Figure \ref{fig:lambda1.0_nu0.1_t5}. 

The dam break wave and the smooth wave have collided and the result of the collision is located approximately in the subdomain \([-3,3]\). The PBC coupling and the PCIF coupling yield different domain decompositions, with the PBC coupling generally yielding higher order than the PCIF coupling in regions where the two coupling methods result in a different order, see Figure \ref{fig:lambda1.0_nu0.1_t5_h_full}. This is observed in the subdomain \([4,13]\), for example. The reason for this is that in the PBC coupling, the cells are grouped into groups of four and given the same order, which is the highest order appearing in the four cells. Moreover, the padded moment in the padded buffer cell in the PBC coupling are in general not equal to zero, such that the first condition in \eqref{eq:residual_decrease_criteria} is not automatically satisfied, contrary to in the PCIF coupling. This is also observed for the other friction cases presented in this section. The differences in both the height \(h\) (Figure \ref{fig:lambda1.0_nu0.1_t5_h}) and the mean velocity \(u_m\) (Figure \ref{fig:lambda1.0_nu0.1_t5_u_full}) between the higher-order model and the lower-order model are small. This can be explained by the fact that the slip length is large compared to the small viscosity, such that there is little friction with the bottom and little internal friction that creates a complex vertical velocity profile. The A-SWME-PBC and A-SWME-PCIF yield similar results for the height \(h\) and the average velocity \(u_m\) that are close to the higher-order model, especially compared to the lower-order model. This is also observed for the first moment \(\alpha_1\) (Figure \ref{fig:lambda1.0_nu0.1_t5_alpha1_full}), for which the difference between the higher-order model and the lower-order model is considerable. However, the differences in $\alpha_1$ for the different models only have a considerable influence on the reconstruction of the vertical velocity profile, but not on the height \(h\) and the mean velocity \(u_m\), such that the \(\text{SWME}_1\) could be used in the entire domain if the main interest is in the height \(h\) and the mean velocity \(u_m\). Note that for the first moment \(\alpha_1\), there is a small region, approximately the interval \([6,7]\) in which the PBC coupling yields a different numerical solution than the PCIF coupling, with the former being closer to the higher-order model. This is a result of the different domain decompositions of the two coupling methods in this interval, see Figure \ref{fig:lambda1.0_nu0.1_t5_h_full}.

Note further that the A-SWME-PBC and the A-SWME-PCIF deviate from the higher-order \(\text{SWME}_5\) for the first moment \(\alpha_1\) at the boundaries, see Figure \ref{fig:lambda1.0_nu0.1_t5_alpha1_full}. Instead, the A-SWME-PBC and the A-SWME-PCIF are close to the lower-order \(\text{SWME}_1\). The same behavior is observed for the following friction cases. Since there are no gradients at the boundaries, the model error at the boundaries can be reduced by reducing the tolerance values \(\epsilon_w\) and \(\kappa_S\) in Table \ref{tab:tolerances}.

\begin{figure}[htb!]
	\centering
	\begin{subfigure}[b]{0.475\textwidth}
		\centering
		\includegraphics[width=\textwidth]{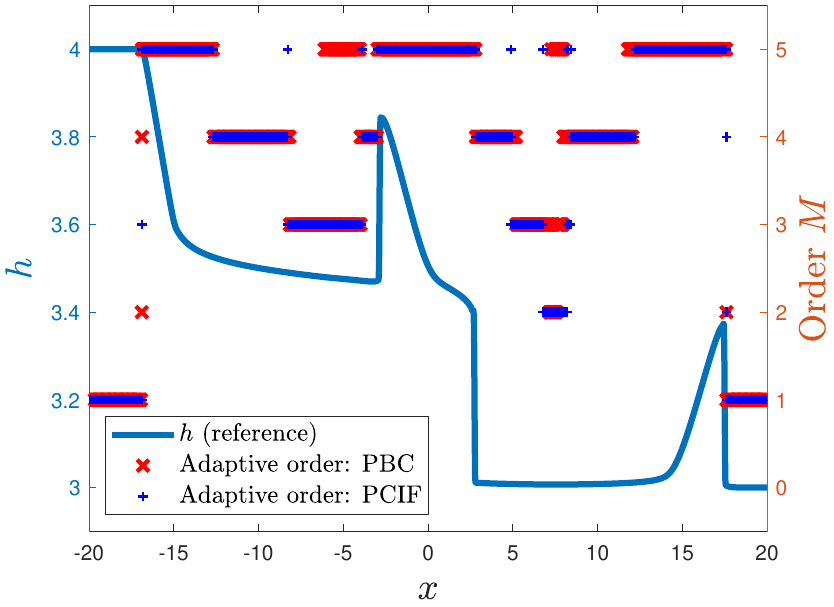}
		\caption{Water height \(h\) alongside adaptive orders}
		\label{fig:lambda0.1_nu0.1_t5_h_full}
	\end{subfigure}
	\hfill
	\begin{subfigure}[b]{0.45\textwidth}
		\centering
		\includegraphics[width=\textwidth]{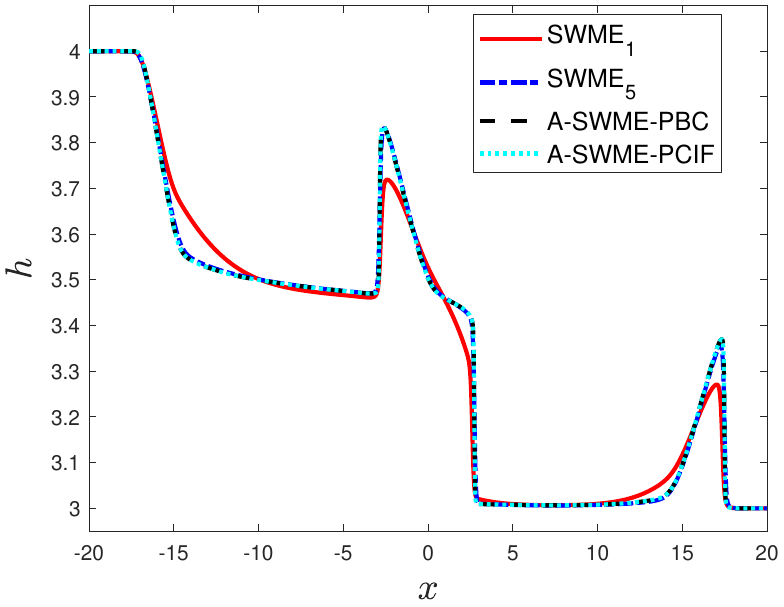}
		\caption{Water height \(h\)}
		\label{fig:lambda0.1_nu0.1_t5_h}
	\end{subfigure}
	\hfill
	\begin{subfigure}[b]{0.45\textwidth}
		\centering
		\includegraphics[width=\textwidth]{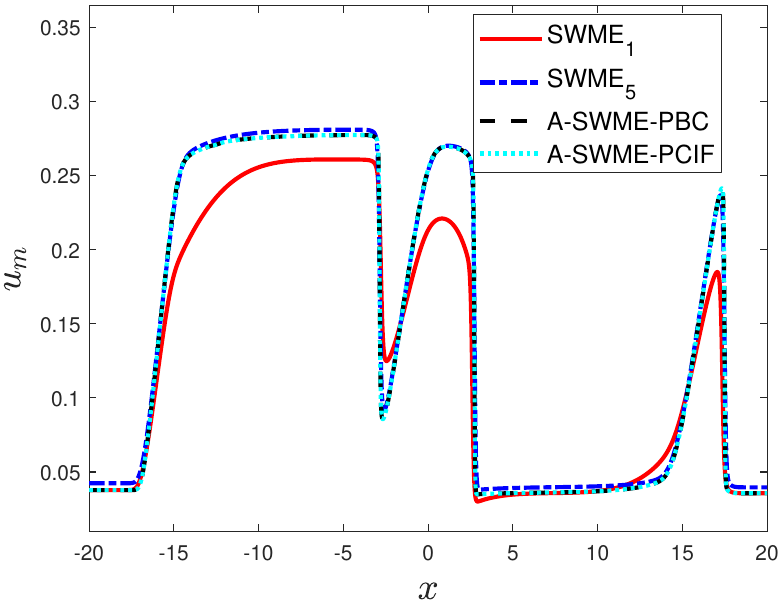}
		\caption{Mean velocity \(u_m\)}
		\label{fig:lambda0.1_nu0.1_t5_u_full}
	\end{subfigure}
	\hfill
	\begin{subfigure}[b]{0.45\textwidth}
		\centering
		\includegraphics[width=\textwidth]{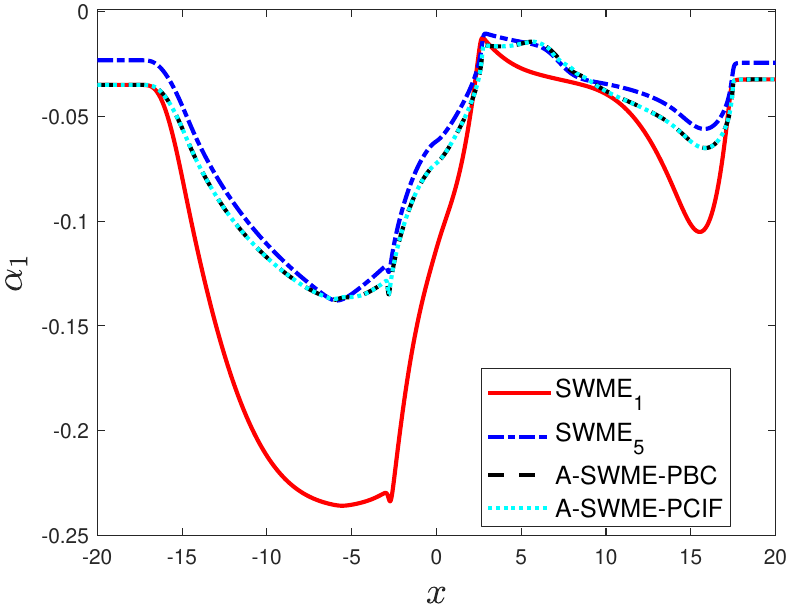}
		\caption{First moment \(\alpha_1\)}
		\label{fig:lambda0.1_nu0.1_t5_alpha1_full}
	\end{subfigure}
	\caption{Collision of a dam break wave with a smooth wave with friction parameters \(\lambda=0.1\) and \(\nu=0.1\) (friction case 2) at time \(t=5\). (a): Solid blue line is the simulated water height from the reference solution. The red crosses and the blue plus signs are the orders in each cell using the PBC coupling and the PCIF coupling, respectively. (b)-(d): The lower-order model is the \(\text{SWME}_1\) (solid red line). The higher-order model is the \(\text{SWME}_5\) (blue dash-dotted line). The A-SWME-PBC (black dashed line) and the A-SWME-PCIF (cyan dotted line) yield similar results, close to the higher-order model, for the height \(h\), mean velocity \(u_m\), and first moment \(\alpha_1\).}
	\label{fig:lambda0.1_nu0.1_t5}
\end{figure}

\paragraph{Friction case 2: \(\lambda = 0.1\) and \(\nu = 0.1\)}
In this friction case, the slip length \(\lambda\) is smaller compared to the previous friction case, so that there is more friction with the bottom, which creates a more complex vertical velocity profile. The viscosity \(\nu\) is small such that there is little internal friction. The adaptive models resulted in a minimum order \(M=1\), so the \(\text{SWME}_1\) was chosen as lower-order model for fair comparison. The results at time \(t=5\) are shown in Figure \ref{fig:lambda0.1_nu0.1_t5}. The water height in the full domain at time \(t=5\) together with the adaptive orders using the PBC coupling (red crosses) and the PCIF coupling (blue plus signs) is shown in Figure \ref{fig:lambda0.1_nu0.1_t5_h_full}.   

The dam break wave and the smooth wave collided similarly as in the previous friction case. Compared to the previous friction case in which the slip length was larger than the viscosity, the differences in both the height \(h\) (Figure \ref{fig:lambda0.1_nu0.1_t5_h}) and the mean velocity \(u_m\) (Figure \ref{fig:lambda0.1_nu0.1_t5_u_full}) between the higher-order model and the lower-order model are considerably larger. The reason for this is that there is more bottom friction, causing higher-order vertical velocity profile deviations. The A-SWME-PBC and A-SWME-PCIF yield nearly identical results for the height \(h\) and the average velocity \(u_m\) that are close to the higher-order model, especially compared to the lower-order model. This is also observed for the first moment \(\alpha_1\) (Figure \ref{fig:lambda0.1_nu0.1_t5_alpha1_full}). 

\begin{figure}[htb!]
	\centering
	\begin{subfigure}[b]{0.475\textwidth}
		\centering
		\includegraphics[width=\textwidth]{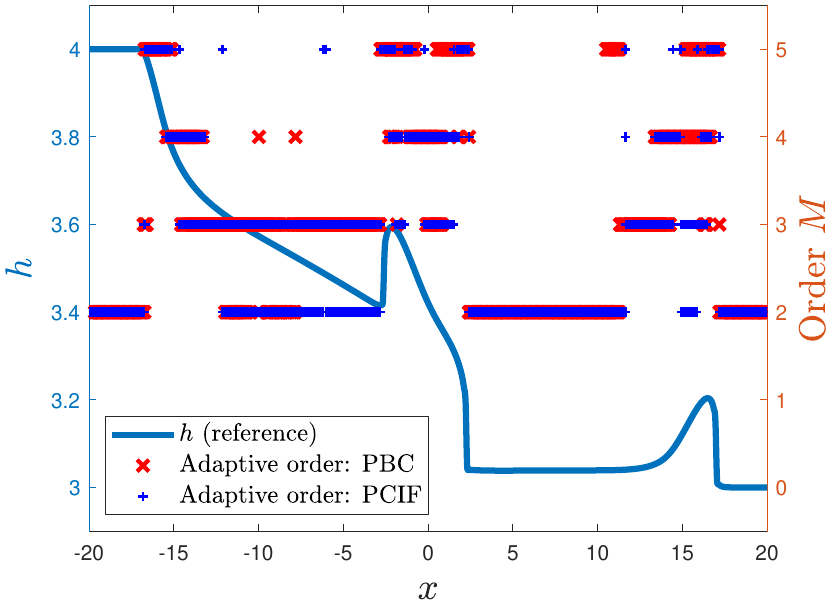}
		\caption{Water height \(h\) alongside adaptive orders}
		\label{fig:lambda0.1_nu1.0_t5_h_full}
	\end{subfigure}
	\hfill
	\begin{subfigure}{0.425\textwidth}
		\centering
		\includegraphics[width=\textwidth]{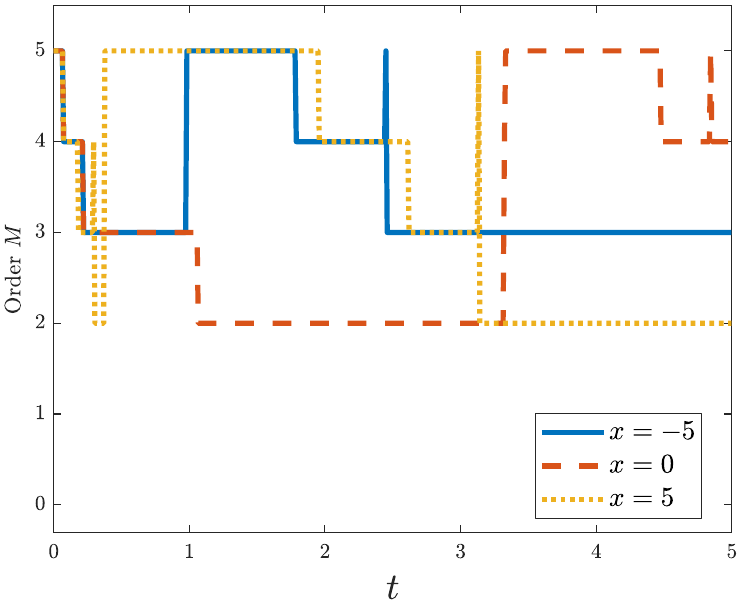}
		\caption{Time evolution of order for PCIF at $x\in\{-5,0,5\}$}
		\label{fig:lambda0.1_nu1.0_t5_momentsEvolution}
	\end{subfigure}
	\hfill
	\begin{subfigure}[b]{0.45\textwidth}
		\centering
		\includegraphics[width=\textwidth]{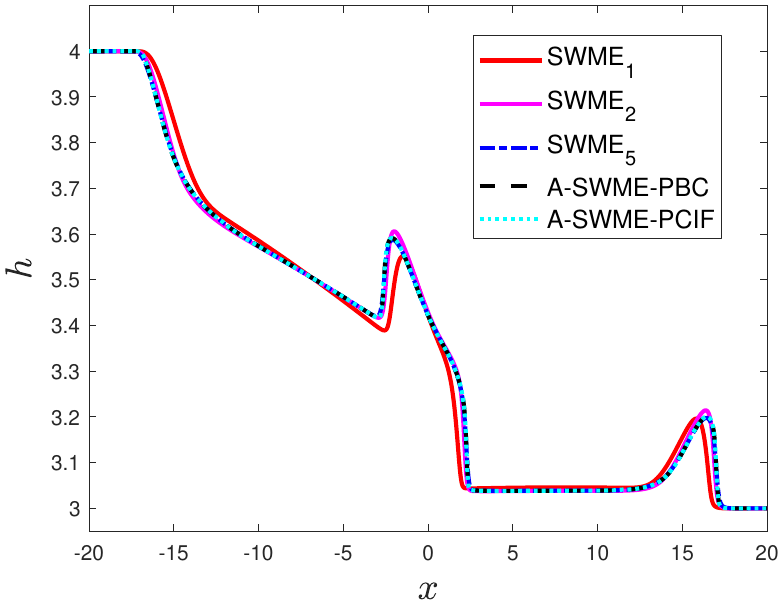}
		\caption{Water height \(h\)}
		\label{fig:lambda0.1_nu1.0_t5_h}
	\end{subfigure}
	\hfill
	\begin{subfigure}[b]{0.45\textwidth}
		\centering
		\includegraphics[width=\textwidth]{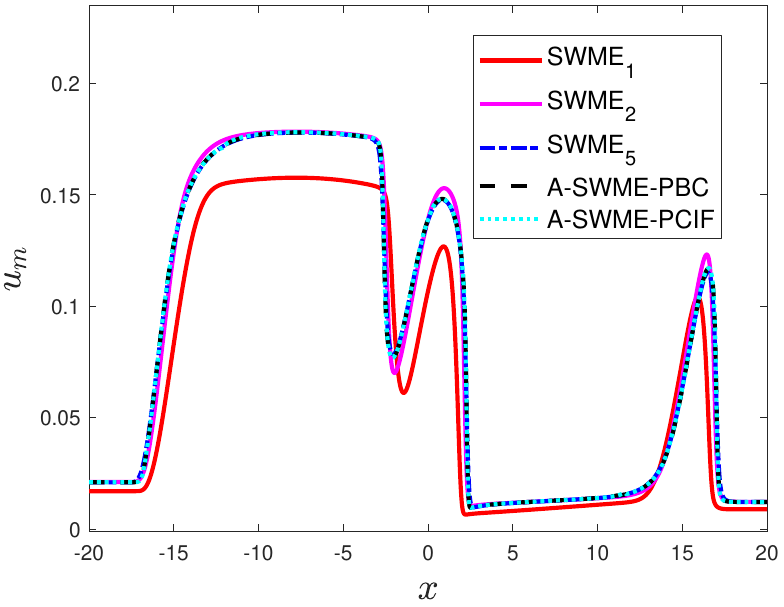}
		\caption{Mean velocity \(u_m\)}
		\label{fig:lambda0.1_nu1.0_t5_um_full}
	\end{subfigure}
	\hfill
	\begin{subfigure}[b]{0.45\textwidth}
		\centering
		\includegraphics[width=\textwidth]{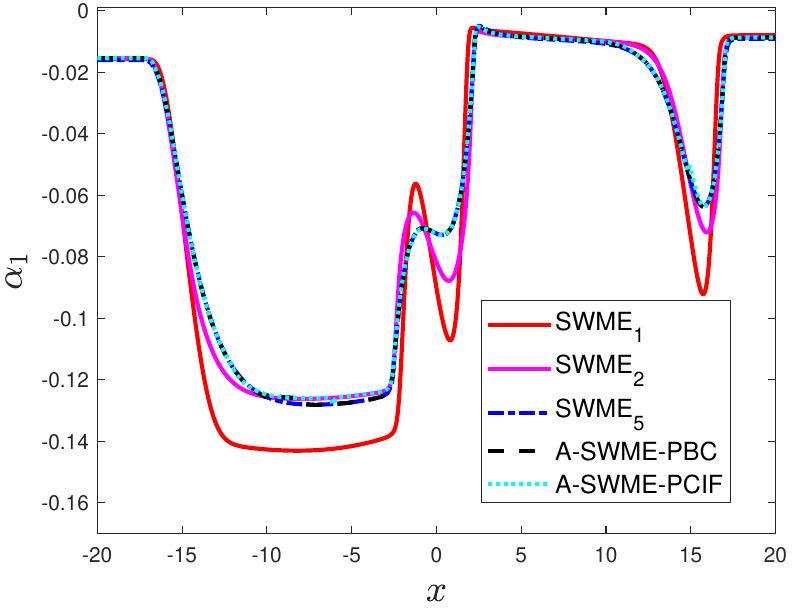}
		\caption{First moment \(\alpha_1\)}
		\label{fig:lambda0.1_nu1.0_t5_alpha1_full}
	\end{subfigure}
	\hfill
	\begin{subfigure}[b]{0.45\textwidth}
		\centering
		\includegraphics[width=\textwidth]{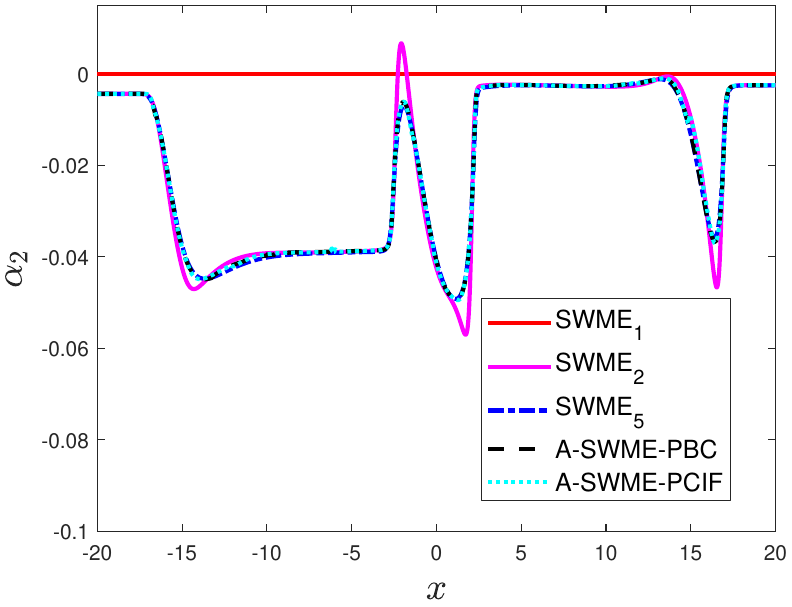}
		\caption{Second moment \(\alpha_2\)}
		\label{fig:lambda0.1_nu1.0_t5_alpha2_full}
	\end{subfigure}
	\caption{Collision of a dam break wave with a smooth wave with friction parameters \(\lambda=0.1\) and \(\nu=1\) (friction case 3) at time \(t=5\). (a): Solid blue line is the simulated water height from the reference solution. The red crosses and the blue plus signs are the orders in each cell using the PBC coupling and the PCIF coupling, respectively. (b): Time-evolution of the adaptive order $M$ at positions $x=-5$ (blue solid line), $x=0$ (red dashed line), and $x=5$ (yellow dotted line). (c)-(f): The lower-order models are the \(\text{SWME}_1\) (solid red line) and the the \(\text{SWME}_2\) (solid magenta line). The higher-order model is the \(\text{SWME}_5\) (blue dash-dotted line). The A-SWME-PBC (black dashed line) and the A-SWME-PCIF (cyan dotted line) yield similar results, close to the higher-order model, for the height \(h\), mean velocity \(u_m\), first moment \(\alpha_1\) and second moment \(\alpha_2\).}
	\label{fig:lambda0.1_nu1.0_t5}
\end{figure}

\paragraph{Friction case 3: \(\lambda = 0.1\) and \(\nu = 1\)}
In the third friction case, the viscosity \(\nu\) is larger compared to the previous friction cases, so that there is more internal friction that can result in a more complex velocity profile. The slip length \(\lambda\) is small, so that there is considerable friction with the bottom, as in the previous friction case. The adaptive models resulted in a minimum order \(M=2\), so the \(\text{SWME}_2\) was chosen as the lower-order model for fair comparison. The results at time \(t=5\) are shown in Figure \ref{fig:lambda0.1_nu1.0_t5}. The results of the \(\text{SWME}_1\) are also shown to justify the necessity of a higher-order moment model. It is observed that the \(\text{SWME}_1\) yielded significantly different numerical results. The water height in the full domain at time \(t=5\) together with the adaptive orders using the PBC coupling (red crosses) and the PCIF coupling (blue plus signs) is shown in Figure \ref{fig:lambda0.1_nu1.0_t5_h_full}. 

The dam break wave and the smooth wave collided similarly as in the previous friction cases. The differences in both the height \(h\) (Figure \ref{fig:lambda0.1_nu1.0_t5_h}) and the mean velocity \(u_m\) (Figure \ref{fig:lambda0.1_nu1.0_t5_um_full}) between the higher-order model and the lower-order model are considerable. The reason for this is that there is large bottom friction and large internal friction, causing higher-order vertical velocity profile deviations. The A-SWME-PBC and A-SWME-PCIF yield similar results for the height \(h\) and the average velocity \(u_m\) that are close to the higher-order model, especially compared to the lower-order model. This is also observed for the first moment \(\alpha_1\) (Figure \ref{fig:lambda0.1_nu1.0_t5_alpha1_full}), apart from in the approximate subdomain \([-10,-6]\), in which the PBC coupling yields results that are closer to the higher-order model, compared to the PCIF coupling. This is a result of the different domain decompositions obtained by the PBC coupling and the PCIF coupling, see Figure \ref{fig:lambda0.1_nu1.0_t5_h_full}. Finally, the A-SWME-PBC and A-SWME-PCIF yield nearly identical results for the second moment \(\alpha_2\) (Figure \ref{fig:lambda0.1_nu1.0_t5_alpha2_full}), apart from in a small region around \(x=15\), in which the PBC coupling yields results that are closer to the higher-order model, compared to the PCIF coupling. Again, this is a result of the different domain decompositions obtained by the PBC coupling and the PCIF coupling. 

Figure \ref{fig:lambda0.1_nu1.0_t5_momentsEvolution} shows the evolution of the adaptive order $M$ at three $x-$positions. At $x=-5$, the order $M$ first decreases to $M=3$, until the shock passes through $x=-5$ at $t\approx 1$ and the maximum order $M_{\text{max}}=5$ is assigned. At $x=0$, the order $M$ gradually decreases to $M=2$, until the shock passes through $x=0$ and interacts with the smooth wave at $t\approx 3.3$, and the maximum order $M_{\text{max}}=5$ is assigned. At $x=5$, the order $M_{\text{max}}$ is assigned as soon as the smooth wave passes through $x=5$ at $t\approx 0.4$, but decreases to $M=2$ at $t\approx 3.2$, when the smooth wave has left the region around $x=5$. This shows that the adaptive SWME is able to dynamically detect shocks and refine and coarsen the model as desired.

In all three test cases, the adaptive models A-SWME-PBC and A-SWME-PCIF yield similar and accurate results, compared to the higher-order \(\text{SWME}_5\). This indicates that the A-SWME-PBC and the A-SWME-PCIF correctly capture the regions that require a higher order.

\subsubsection{Runtime and error comparison}
The model-adaptive simulation of the SWME aims at obtaining a computational speedup while achieving a certain accuracy. Therefore, the runtimes and the error norms with respect to a reference solution are compared here for the test case described in Table \ref{tab:setup-general}. 
\paragraph{Runtime comparison}
The relative runtime of the simulations run with the different models is shown in Table \ref{tab:simulation_time}. 
\begin{table}[h!]
	\centering
	\small
	\setlength{\tabcolsep}{5pt}
	\renewcommand{\arraystretch}{1.15}
	\begin{tabular}{l
        S[table-format=1.2, scientific-notation=false]
        S[table-format=1.2, scientific-notation=false]
        S[table-format=1.2, scientific-notation=false]
        S[table-format=1.2, scientific-notation=false]}
		\toprule
		\multirow{2}{*}{\makecell[c]{Friction\\\(\lambda,\nu\)}} &
		\multicolumn{4}{c}{Relative runtime} \\
		\cmidrule(lr){2-5}
		& \multicolumn{1}{c}{\makecell[c]{Lower-order\\\(\mathrm{SWME}_1/\mathrm{SWME}_2\)}} &
		\multicolumn{1}{c}{\makecell[c]{Higher-order\\\(\mathrm{SWME}_5\)}} &
		\multicolumn{1}{c}{\makecell[c]{Adaptive\\A-SWME-PBC}} &
		\multicolumn{1}{c}{\makecell[c]{Adaptive\\A-SWME-PCIF}} \\
		\midrule
		\(\lambda=1,\ \nu=0.1\)   & 0.27 & 1.00 & 0.70 & 0.62 \\
		\(\lambda=0.1,\ \nu=0.1\) & 0.24 & 1.00 & 0.70 & 0.63 \\
		\(\lambda=0.1,\ \nu=1\)   & 0.34 & 1.00 & 0.66 & 0.73 \\
		\bottomrule
	\end{tabular}
	\caption{\label{tab:simulation_time}Relative runtime comparison of the lower-order \(\text{SWME}_1\) (for \(\lambda=1,\nu=0.1\) and \(\lambda=0.1,\nu=0.1\)) or \(\text{SWME}_2\) (for \(\lambda=0.1,\nu=1\)), the higher-order \(\text{SWME}_5\), A-SWME-PBC and the A-SWME-PCIF for the test case described in Table \ref{tab:setup-general}, relative to the runtime of the higher-order \(\text{SWME}_5\). The adaptive models A-SWME-PBC and A-SWME-PCIF yield speedups of up to 38 percent compared to the \(\text{SWME}_5\).}
\end{table}
For each friction case, there is a significant decrease in runtime of up to 38 percent using the A-SWME-PBC or A-SWME-PCIF compared to the higher-order model \(\text{SWME}_5\), with the A-SWME-PBC having a slightly smaller runtime than the A-SWME-PCIF. This speedup already includes the evaluation of the domain decomposition criteria, which therefore yield only small overhead.

\paragraph{Error norm comparison}
The error norms of the numerical solution with respect to the reference solution are shown in Table \ref{tab:error} for the different friction cases. The reference solution and the moment models were each simulated on successively finer spatial and temporal grids until grid convergence was achieved for each method separately. For each model
\(m \in \{\mathrm{SWME}_1/\mathrm{SWME}_2,\mathrm{SWME}_5,\mathrm{A\text{-}SWME\text{-}PBC},\mathrm{A\text{-}SWME\text{-}PCIF}\},\)
and for each quantity \(q \in \{h,u_m,\alpha_1\}\) (and \(q=\alpha_2\) in friction case 3), we compute the relative error with respect to the reference solution \(q_{\mathrm{ref}}\) using the discrete \(L^2\)-norm as
\begin{equation}\label{eq:error_computation}
	\mathrm{Err}(q) \;=\; \frac{\|q-q_{\mathrm{ref}}\|_2}{\|q_{\mathrm{ref}}\|_2}.
\end{equation}

For the friction case \(\lambda=1\) and \(\nu=0.1\), it is observed in Figure \ref{fig:lambda1.0_nu0.1_t5} that the lower-order SWME model is already accurate for the height \(h\) and the mean velocity \(u_m\), explaining why the error norms for the different models are of the same order in this friction case. The adaptive models A-SWME-PBC and A-SWME-PCIF result in smaller error norms than the higher-order \(\text{SWME}_5\), which is likely a result of the numerical error that is of the same order as the model error for the height \(h\) and the mean velocity \(u_m\) in this friction case. This is also observed for the other friction cases in Table \ref{tab:error}. For the first-order moment \(\alpha_1\), the error norms of the A-SWME-PBC and A-SWME-PCIF are approximately one order of magnitude larger than the error norm of the higher-order SWME model, whereas the error norm of the lower-order SWME model is approximately two orders of magnitude larger than the error norm of the higher-order SWME model. 

For the friction case \(\lambda=\nu=0.1\), the error norms of the A-SWME-PBC and A-SWME-PCIF are approximately one order of magnitude smaller than the error norm of the lower-order SWME model for all the variables, and of the same order of magnitude as the higher-order SWME model for the water height $h$ and the mean velocity $u_m$, and one order of magnitude larger than the higher-order SWME model for \(\alpha_1\). 

Finally, for the friction case \(\lambda=0.1\) and \(\nu=1\), the error norms of the A-SWME-PBC and A-SWME-PCIF are of the same order as the higher-order SWME model and between 3 and 8 times smaller than the error norms of the lower-order SWME model for all the variables.

\begin{table}[h!]
	\centering
	\small
	\setlength{\tabcolsep}{5pt}
	\renewcommand{\arraystretch}{1.15}
	\begin{tabular}{
			ll
			S[table-format=1.1e1]
			S[table-format=1.1e1]
			S[table-format=1.1e1]
			S[table-format=1.1e1]
		}
		\toprule
		\multirow{2}{*}{\makecell[c]{Friction\\\(\lambda,\nu\)}} &
		\multirow{2}{*}{Var.} &
		\multicolumn{4}{c}{Error norm} \\
		\cmidrule(lr){3-6}
		& &
		\multicolumn{1}{c}{\makecell[c]{Lower-order\\\(\mathrm{SWME}_1/\mathrm{SWME}_2\)}} &
		\multicolumn{1}{c}{\makecell[c]{Higher-order\\\(\mathrm{SWME}_5\)}} &
		\multicolumn{1}{c}{\makecell[c]{Adaptive\\A-SWME-PBC}} &
		\multicolumn{1}{c}{\makecell[c]{Adaptive\\A-SWME-PCIF}} \\
		\midrule
		
		\multirow{3}{*}{\makecell[c]{\(\lambda = 1\)\\\(\nu = 0.1\)}} 
		& \(h\)        & 3.5e-3 & 3.3e-3 & 1.4e-3 & 1.4e-3 \\
		& \(u_m\)      & 3.4e-2 & 2.9e-2 & 1.3e-2 & 1.3e-2 \\
		& \(\alpha_1\)  & 3.1e-1 & 5.0e-3 & 4.6e-2 & 4.6e-2 \\
		\midrule
		
		\multirow{3}{*}{\makecell[c]{\(\lambda = 0.1\)\\\(\nu = 0.1\)}} 
		& \(h\)        & 1.3e-2 & 3.0e-3 & 1.8e-3 & 1.8e-3 \\
		& \(u_m\)      & 1.6e-1 & 3.0e-2 & 2.3e-2 & 2.3e-2 \\
		& \(\alpha_1\)  & 8.2e-1 & 8.0e-3 & 1.1e-1 & 1.1e-1 \\
		\midrule
		
		\multirow{4}{*}{\makecell[c]{\(\lambda = 0.1\)\\\(\nu = 1\)}} 
		& \(h\)        & 3.1e-3 & 1.7e-3 & 1.1e-3 & 1.1e-3 \\
		& \(u_m\)      & 5.4e-2 & 2.9e-2 & 1.8e-2 & 1.8e-2 \\
		& \(\alpha_1\)  & 6.6e-2 & 9.0e-3 & 7.9e-3 & 1.4e-2 \\
		& \(\alpha_2\)  & 9.0e-2 & 2.9e-2 & 2.0e-2 & 2.9e-2 \\
		\bottomrule
	\end{tabular}
	\caption{\label{tab:error}Error norm comparison of the lower-order \(\text{SWME}_1\), the higher-order \(\text{SWME}_5\), the A-SWME-PBC and A-SWME-PCIF for the three test cases described in Table \ref{tab:setup-general}. The adaptive models A-SWME-PBC and A-SWME-PCIF yield comparable accuracy as the \(\text{SWME}_5\).}
\end{table}

The runtime and error comparison shows the decrease in runtime of the A-SWME-PBC and the A-SWME-PCIF compared to the higher-order SWME model while simultaneously keeping the accuracy close. Furthermore, the A-SWME-PBC and the A-SWME-PCIF yield similar domain decompositions, similar results and similar runtimes. 

\subsection{Numerical results: dam break}
As a second test case, we consider a dam break with initial conditions
\begin{equation}\label{eq:initialConditions_damBreakLarge}
	h(0,x) = 
	\begin{cases}
		5, &  x \leq 0 \\
		1, &  x > 0
	\end{cases}, \qquad u_m(0,x) = 0, \qquad \alpha_i(0,x) = 0\; (i\geq 1).
\end{equation}
The friction parameters are set to \(\nu=1\), \(\lambda = 0.1\) (Friction case 3 in the previous section). For the moment model simulations, the CFL number was set to 0.5 and the grid cell size to \(\Delta x = 0.001\). For the reference solution, outlined in Section \ref{section:referenceSolution}, the grid cell size was set to \(\Delta x_{\text{ref}}=0.00625\) in \(x-\)direction and \(\Delta \zeta_{\text{ref}} = 0.025\) in \(\zeta -\)direction, and the CFL number was set to 0.5.

\begin{figure}[htb!]
	\centering
	\begin{subfigure}[b]{0.45\textwidth}
		\centering
		\includegraphics[width=\textwidth]{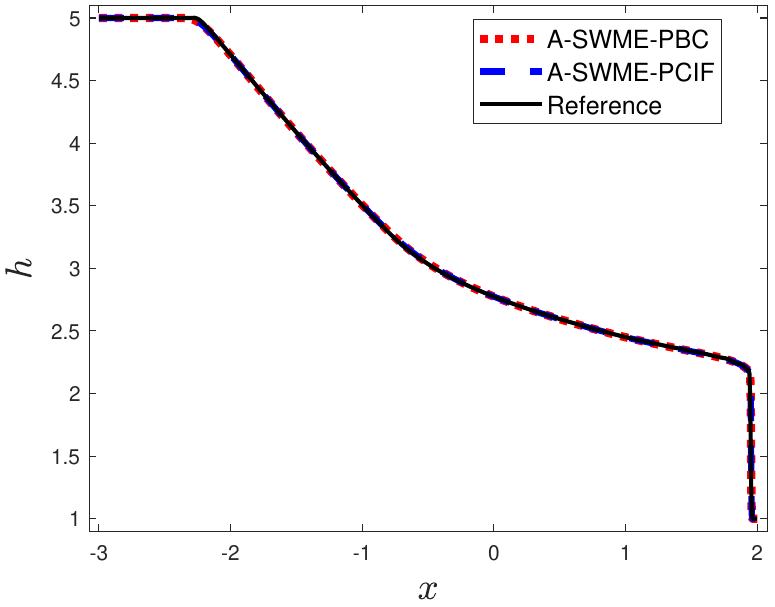}
		\caption{Water height \(h\)}
		\label{fig:damBreakLarge_lambda0.1_nu1.0_t1_h}
	\end{subfigure}
	\hfill
	\begin{subfigure}{0.45\textwidth}
		\centering
		\includegraphics[width=\textwidth]{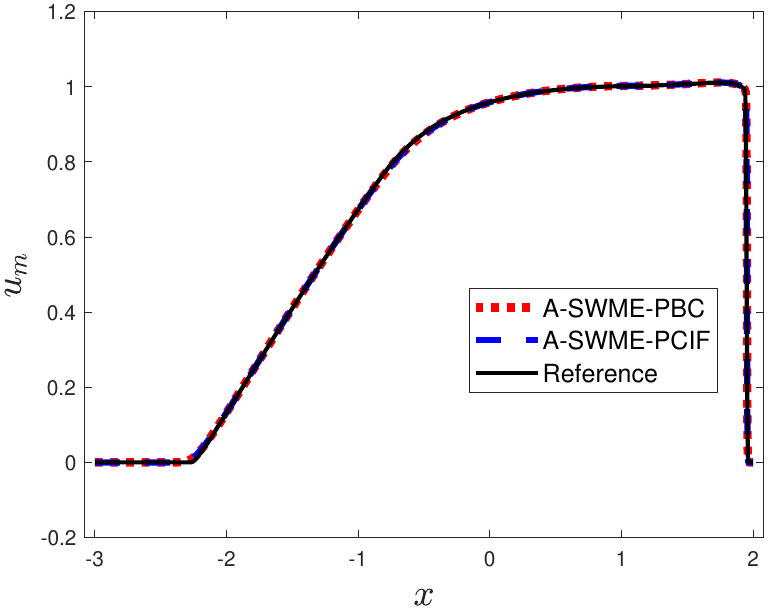}
		\caption{Mean velocity $u_m$}
		\label{fig:damBreakLarge_lambda0.1_nu1.0_t1_u}
	\end{subfigure}
	\caption{Dam break with friction parameters \(\lambda=0.1\) and \(\nu=1\) at time \(t=1\). (a): Water height $h$. (b): Mean velocity $u_m$. The results of the A-SWME-PBC (red dotted line), the A-SWME-PCIF (blue dashed line), and the reference solution (black solid line) are shown for the water $h$ (panel (a)) and the Mean velocity $u_m$ (panel (b)). The A-SWME-PBC and the A-SWME-PCIF yield similar results, close to the reference solution, for the height \(h\) and the mean velocity \(u_m\).}
	\label{fig:damBreakLarge_lambda0.1_nu1.0_t1}
\end{figure}
The results are shown in Figure \ref{fig:damBreakLarge_lambda0.1_nu1.0_t1}. Both for the water height $h$ (Figure \ref{fig:damBreakLarge_lambda0.1_nu1.0_t1_h}) and the mean velocity $u_m$ (Figure \ref{fig:damBreakLarge_lambda0.1_nu1.0_t1_u}), the A-SWME-PBC and the A-SWME-PCIF yield accurate results, compared to the reference solution, and the solutions cannot be visibly distinguished. During the simulation, the orders in subdomains away from the dam-break waves were dynamically set to zero by the domain decomposition criteria, until the dam-break waves passed through the subdomain, after which the order was rapidly increased to the maximum order \(M_{\text{max}}=5\). Because of the large gradients induced by the dam break and because of the large viscosity $\nu$ and small slip length $\lambda$, the maximum order \(M_{\text{max}}=5\) was assigned to nearly every cell that was reached by the dam-break waves and the order was not reduced for the rest of the simulation.  Therefore, the adaptive orders are not shown. Again, the adaptive methods A-SWME-PBC and A-SWME-PCIF are able to accurately track the region that requires refinement and yield accurate results.
\subsubsection{Runtime and error comparison}
The relative runtimes, relative to the lowest-order SWME$_0$, and the relative error norms, relative to the reference solution, of the fixed-order SWME models and of the A-SWME-PBC and the PCIF-SWME are given in Table \ref{tab:runtime_damBreakLarge} and Table \ref{tab:errors_damBreakLarge}, respectively. 
\begin{table}[t]
	\centering
	\small
	\setlength{\tabcolsep}{6pt}
	\begin{tabular}{l
			S[table-format=2.2]
			S[table-format=2.2]
			S[table-format=2.2]
			S[table-format=2.2]
			S[table-format=2.2]
			S[table-format=2.2]
			S[table-format=2.2]
			S[table-format=2.2]}
		\toprule
		& {SWME$_0$} & {SWME$_1$} & {SWME$_2$} & {SWME$_3$} & {SWME$_4$} & {SWME$_5$} & {A-SWME-PBC} & {A-SWME-PCIF} \\
		\midrule
		Relative runtime &
		\multicolumn{1}{c}{0.19} &
		\multicolumn{1}{c}{0.25} &
		\multicolumn{1}{c}{0.34} &
		\multicolumn{1}{c}{0.47} &
		\multicolumn{1}{c}{0.70} &
		\multicolumn{1}{c}{1.00} &
		\multicolumn{1}{c}{0.60} &
		\multicolumn{1}{c}{0.60} \\
		\bottomrule
	\end{tabular}
	\caption{Relative runtime comparison of the models $\text{SWME}_0,\ldots,\text{SWME}_5$, A-SWME-PBC, and A-SWME-PCIF, for dam break test case with friction parameters \(\nu=1\) and \(\lambda = 0.1\), relative to the runtime of the highest-order \(\text{SWME}_5\). The adaptive models A-SWME-PBC and A-SWME-PCIF reduce the runtime by 40 percent compared to \(\text{SWME}_5\).}
\label{tab:runtime_damBreakLarge}
\end{table}

\begin{table}[t]
	\centering
	\small
	\setlength{\tabcolsep}{4pt}
	\begin{tabular}{l
			S[table-format=1.2e-1]
			S[table-format=1.2e-1]
			S[table-format=1.2e-1]
			S[table-format=1.2e-1]
			S[table-format=1.2e-1]
			S[table-format=1.2e-1]
			S[table-format=1.2e-1]}
		\toprule
		Model & {$h$} & {$u_m$} & {$\alpha_1$} & {$\alpha_2$} & {$\alpha_3$} & {$\alpha_4$} & {$\alpha_5$} \\
		\midrule
		SWME$_0$   & \num{1.6e-1} &  \num{6.9e-1} & \multicolumn{1}{c}{--} & \multicolumn{1}{c}{--} & \multicolumn{1}{c}{--} & \multicolumn{1}{c}{--} & \multicolumn{1}{c}{--}\\
		SWME$_1$   & \num{4.0e-2} &\num{1.5e-1} & \num{4.3e-1} & \multicolumn{1}{c}{--} & \multicolumn{1}{c}{--} & \multicolumn{1}{c}{--}& \multicolumn{1}{c}{--} \\
		SWME$_2$   & \num{8.3e-3} &  \num{2.2e-2} & \num{1.3e-1} & \num{1.9e-1} & \multicolumn{1}{c}{--} & \multicolumn{1}{c}{--} & \multicolumn{1}{c}{--} \\
		SWME$_3$   & \num{3.8e-3} &  \num{1.3e-2} & \num{1.9e-2} & \num{1.0e-1} & \num{1.8e-1} & \multicolumn{1}{c}{--}& \multicolumn{1}{c}{--} \\
		SWME$_4$   & \num{3.6e-3} & \num{1.3e-2} & \num{1.5e-2} & \num{2.8e-2} & \num{1.3e-1} & \num{1.9e-1} & \multicolumn{1}{c}{--} \\
		SWME$_5$   & \num{3.6e-3} & \num{1.3e-2} & \num{1.5e-2} & \num{2.5e-2} & \num{6.0e-2} & \num{1.6e-1} & \num{2.1e-1} \\
		A-SWME-PBC & \num{3.6e-3} & \num{1.3e-2} & \num{1.5e-2} & \num{2.5e-2} & \num{6.0e-2} & \num{1.6e-1} & \num{2.1e-1} \\
		A-SWME-PCIF & \num{3.6e-3} & \num{1.3e-2} & \num{1.5e-2} & \num{2.5e-2} & \num{6.0e-2} & \num{1.6e-1} & \num{2.1e-1} \\
		\bottomrule
	\end{tabular}
	\caption{Error norm comparison for $h$, $u_m$, $\alpha_i$ ($i=1,\ldots,5$), across the models $\text{SWME}_0,\ldots,\text{SWME}_5$, A-SWME-PBC, and A-SWME-PCIF, for dam break test case with friction parameters \(\nu=1\) and \(\lambda = 0.1\). The adaptive models A-SWME-PBC and A-SWME-PCIF yield accurate results for all variables, comparable to the accuracy of the highest-order $\text{SWME}_5$.}
\label{tab:errors_damBreakLarge}
\end{table}

For the water height $h$ and the mean velocity $u_m$, the $\text{SWME}_3$ already yields accurate results and can also be used as the highest-order SWME model in the adaptive simulations. If the interest is in higher-order moments and reconstructing the vertical velocity profile of the horizontal velocity, higher-order SWME models are required for accurate results. It is observed that the accuracy of the adaptive models A-SWME-PBC and A-SWME-PCIF is comparable to the accuracy of the highest-order \(\text{SWME}_5\). Moreover, the adaptive models A-SWME-PBC and A-SWME-PCIF yield computational speedups of 40 percent compared to the highest-order $\text{SWME}_5$.

This dam break test case shows how the A-SWME-PBC and the A-SWME-PCIF automatically increase the order in cells in which high-gradients are detected or in which the vertical velocity profile of the horizontal velocity is strongly non-uniform (as a result of large bottom friction, for example). Moreover, both the PCIF coupling and the PBC coupling yield stable results.  

\section{Conclusion}\label{section:Conclusion}
In this paper, we proposed a space- and time-model-adaptive procedure for the simulation of a free surface flow modeled by the hierarchical SWME. The adaptive procedure consists of a domain decomposition and a spatial coupling method. The domain decomposition is based on a decomposition of the SWME model that exploits the hierarchical structure of the SWME, resulting in a set of error estimators and a set of domain decomposition criteria. This effectively decomposes the domain into a set of subdomains each modeled by a SWME model of a suitable order. Next, the subdomains are coupled at their boundary interfaces using a spatial coupling method. We proposed two different spatial coupling methods: (1) The Padded Buffer Cell (PBC) coupling uses a padded buffer cell with dynamically updated padded values, allowing flexibility in the treatment of artificial jumps in the higher-order moments at the boundary interfaces, but it does not admit a path-conservative formulation; (2) the Path-Conservative Interface Flux (PCIF) coupling sets the padded values to zero, yielding a path-conservative coupling method that reduces to conservation form for conservative moment equations, but includes the artificial jump contributions in the higher-order moments, which may create small oscillations. Time adaptivity is included by decomposing the domain and accordingly applying the spatial coupling formulas in every time step. We performed numerical simulations of a dam break colliding with a smooth wave and of a classical dam break. The adaptive models with PBC coupling and with PCIF coupling showed accurate results and significant speedups of up to \(40\) percent compared to a higher-order model.

Despite these promising results, the adaptive models proposed in this paper remain limited to one-dimensional configurations without bottom-topography effects. Several extensions therefore remain for future research. In particular, incorporating bottom topography would require a well-balanced formulation, while the treatment of time-dependent boundary conditions for moment models is not straightforward. Both features are of clear practical interest and may further broaden the applicability of the proposed adaptive method to more complex flow configurations. Ongoing work focuses on the application of the model-adaptive simulation procedure to two-dimensional flow with more complex flow configurations and to different hierarchical moment models. A suggestion for future work is the combination of the model-adaptive simulation procedure with data-driven methods for the estimation of the model error.

\section{CRediT authorship contribution statement}

\textbf{R. Verbiest:} Conceptualization, Methodology, Software, Formal Analysis, Investigation, Data Curation, Writing - Original Draft, Writing - Review \& Editing, Visualization. \textbf{J. Koellermeier:} Conceptualization, Methodology, Writing - Review \& Editing, Supervision, Funding acquisition.

\section{Acknowledgments}
This work is part of the project \textit{HiWAVE} with file number VI.Vidi.233.066 of the \textit{ENW Vidi} research programme, funded by the \textit{Dutch Research Council (NWO)}.

\appendix

\section{Analytical formulas for \(\boldsymbol{\delta A_{M_{\mathrm{L}},M_{\mathrm{H}}}(\vec{w}_{M_{\mathrm{H}}})}\) and \(\boldsymbol{\delta \vec{S}_{M_{\mathrm{L}},M_{\mathrm{H}}}(\vec{w}_{M_{\mathrm{H}}})}\)}\label{sec:appendix}

Let \(M_{\mathrm{L}},M_{\mathrm{H}}\in\mathbb{N}\) be the orders of two SWME models, with \(M_{\mathrm{L}}<M_{\mathrm{H}}\). The appendix uses the notation introduced in Section \ref{sec:domain-decomposition}. Recall the definition of the system matrix of the \(\text{SWME}_{M_{\mathrm{H}}}\):
\begin{equation}\label{eq:systemMatr-x_Mh_appendix}
	A_{M_{\mathrm{H}}}(\vec{w}_{M_{\mathrm{H}}}) = \frac{\partial \vec{F}_{M_{\mathrm{H}}}(\vec{w}_{M_{\mathrm{H}}})}{\partial \vec{w}_{M_{\mathrm{H}}}}-Q_{M_{\mathrm{H}}}(\vec{w}_{M_{\mathrm{H}}})\in\mathbb{R}^{({M_{\mathrm{H}}}+2)\times({M_{\mathrm{H}}}+2)},
\end{equation}
where \(\vec{F}(\vec{w}_{M_{\mathrm{H}}})\in\mathbb{R}^{M_{\mathrm{H}}+2}\) is the conservative flux and where \(Q_{M_{\mathrm{H}}}(\vec{w}_{M_{\mathrm{H}}})\in\mathbb{R}^{(M_{\mathrm{H}}+2)\times(M_{\mathrm{H}}+2)}\) contains the non-conservative flux terms. The quantity \(\delta A_{M_{\mathrm{L}},M_{\mathrm{H}}}(\vec{w}_{M_{\mathrm{H}}})\in\mathbb{R}^{(M_{\mathrm{L}}+2)\times(M_{\mathrm{L}}+2)}\) defined by Equation \eqref{eq:A-S-decomposition1} is split into a part corresponding to the conservative terms and a part corresponding to the non-conservative terms:
\begin{equation}\label{eq:diffs_split}
	\delta A_{M_{\mathrm{L}},M_{\mathrm{H}}}(\vec{w}_{M_{\mathrm{H}}}) = \frac{\partial \delta \vec{F}_{M_{\mathrm{L}},M_{\mathrm{H}}}(\vec{w}_{M_{\mathrm{H}}})}{\partial \vec{w}_{M_{\mathrm{L}}}} - \delta Q_{M_{\mathrm{L}},M_{\mathrm{H}}}(\vec{w}_{M_{\mathrm{H}}}).
\end{equation}
Here, the conservative part \(\delta \vec{F}_{M_{\mathrm{L}},M_{\mathrm{H}}}(\vec{w}_{M_{\mathrm{H}}})\in\mathbb{R}^{M_{\mathrm{L}}+2}\) is defined as
\begin{equation}\label{eq:diff_cons_split}
	\delta \vec{F}_{M_{\mathrm{L}},M_{\mathrm{H}}}(\vec{w}_{M_{\mathrm{H}}}) := \vec{F}_{:M_{\mathrm{L}}}(\vec{w}_{M_{\mathrm{H}}}) - \vec{F}_{M_{\mathrm{L}}}(\vec{w}_{:M_{\mathrm{L}}})\in\mathbb{R}^{M_{\mathrm{L}}+2},
\end{equation}
where the first \(M_{\mathrm{L}}+2\) entries of \(\vec{F}_{M_{\mathrm{H}}}(\vec{w}_{M_{\mathrm{H}}})\in\mathbb{R}^{M_{\mathrm{H}}+2}\) are defined by \(\vec{F}_{:M_{\mathrm{L}}}(\vec{w}_{M_{\mathrm{H}}})\in\mathbb{R}^{M_{\mathrm{L}}+2}\), and the non-conservative part \(\delta Q_{M_{\mathrm{L}},M_{\mathrm{H}}}(\vec{w}_{M_{\mathrm{H}}})\in\mathbb{R}^{(M_{\mathrm{L}}+2)\times(M_{\mathrm{L}}+2)}\) is defined as
\begin{equation}\label{eq:appendix_deltaQ-def}
	\delta Q_{M_{\mathrm{L}},M_{\mathrm{H}}}(\vec{w}_{M_{\mathrm{H}}}):=Q_{:M_{\mathrm{L}},:M_{\mathrm{L}}}(\vec{w}_{M_{\mathrm{H}}})-Q_{M_{\mathrm{L}}}(\vec{w}_{M_{\mathrm{L}}})\in\mathbb{R}^{(M_{\mathrm{L}}+2)\times(M_{\mathrm{L}}+2)},
\end{equation}
where \(Q_{:M_{\mathrm{L}},:M_{\mathrm{L}}}(\vec{w}_{M_{\mathrm{H}}})\in\mathbb{R}^{(M_{\mathrm{L}}+2)\times(M_{\mathrm{L}}+2)}\) are the first \(M_{\mathrm{L}}+2\) rows and the first \(M_{\mathrm{L}}+2\) columns of the matrix \(Q_{M_{\mathrm{H}}}(\vec{w}_{M_{\mathrm{H}}})\). 

\paragraph{Conservative part.}
First, the Jacobian matrix \(\frac{\partial \delta \vec{F}_{M_{\mathrm{L}},M_{\mathrm{H}}}(\vec{w}_{M_{\mathrm{H}}})}{\partial \vec{w}_{M_{\mathrm{L}}}}\in\mathbb{R}^{(M_{\mathrm{L}}+2)\times(M_{\mathrm{L}}+2)}\) is calculated. Writing
\begin{equation}
	\delta \vec{F}_{M_{\mathrm{L}},M_{\mathrm{H}}}(\vec{w}_{M_{\mathrm{H}}}) = (\delta F_0,\delta F_1,\ldots,\delta F_{M_{\mathrm{L}}+1})^T,
\end{equation}
the vector \(\delta \vec{F}_{M_{\mathrm{L}},M_{\mathrm{H}}}(\vec{w}_{M_{\mathrm{H}}})\) can be constructed explicitly by computing the entries \(\delta F_j\), \(j=0,1,\ldots,M_{\mathrm{L}}+1\), which can be computed by subtracting the conservative flux terms appearing in the mass balance equation \eqref{eq:SWME-h}, the momentum balance equation \eqref{eq:SWME-hum}, and the moment equations for \(\alpha_i\) \eqref{eq:SWME-halphai}, \(i=1,\ldots,M_{\mathrm{L}}\), of the \(\text{SWME}_{M_{\mathrm{L}}}\) from the respective conservative flux terms in the first \(M_{\mathrm{L}}+2\) equations of the \(\text{SWME}_{M_{\mathrm{H}}}\). We obtain
\begin{equation}\label{eq:diff_F0_F1_Fi+1}
		\delta F_0 = 0, \quad\;\; \delta F_1 = h\sum_{k=M_{\mathrm{L}}+1}^{M_{\mathrm{H}}}\frac{\alpha_k^2}{2k+1}, \quad\;\; \delta F_{i+1} = h\left( 2\sum_{j=1}^{M_{\mathrm{L}}}\sum_{k=M_{\mathrm{L}}+1}^{M_{\mathrm{H}}}A_{ijk}\alpha_j\alpha_k+\sum_{j,k=M_{\mathrm{L}}+1}^{M_{\mathrm{H}}}A_{ijk}\alpha_j\alpha_k \right), \, i=1,\ldots,M_{\mathrm{L}}.
\end{equation}
The partial derivatives of \(\delta F_1\) and \(\delta F_{i+1}\), \(i=1,\ldots,M_{\mathrm{L}}\), with respect to \(h,hu_m\) and \(h\alpha_m\), \(m=1,\ldots,M_{\mathrm{L}}\), can then be easily computed using \eqref{eq:diff_F0_F1_Fi+1}. We obtain
\begin{equation}\label{eq:diff_F1_partDer}
	\frac{\partial \delta F_1}{\partial h} = -\sum_{i=M_{\mathrm{L}}+1}^{M_{\mathrm{H}}}\frac{\alpha_i^2}{2i+1}, \quad \frac{\partial \delta F_1}{\partial (hu_m)} = 0, \quad \frac{\partial \delta F_1}{\partial (h\alpha_m)_{m=1,\ldots,M_{\mathrm{L}}}} = 0,
\end{equation}
and
\begin{equation}
		\label{eq:diff_Fi_partDer}
		\frac{\partial \delta F_{i+1}}{\partial h} = -2\sum_{j=1}^{M_{\mathrm{L}}}\sum_{k=M_{\mathrm{L}}+1}^{M_{\mathrm{H}}}A_{ijk}\alpha_j\alpha_k-\sum_{j,k=M_{\mathrm{L}}+1}^{M_{\mathrm{H}}}A_{ijk}\alpha_j\alpha_k, \quad \frac{\partial \delta F_{i+1}}{\partial (hu_m)} = 0, \quad\frac{\partial \delta F_{i+1}}{\partial (h\alpha_m)_{m=1,\ldots,M_{\mathrm{L}}}} = 2\sum_{k=M_{\mathrm{L}}+1}^{M_{\mathrm{H}}}A_{imk}\frac{\alpha_k}{2k+1}.
\end{equation}
The Jacobian matrix \(\frac{\partial \delta \vec{F}_{M_{\mathrm{L}},M_{\mathrm{H}}}(\vec{w}_{M_{\mathrm{H}}})}{\partial \vec{w}_{M_{\mathrm{L}}}}\) is thus given by
\begin{equation}\label{eq:delta_F_full-matrix}
	\frac{\partial \delta \vec{F}_{M_{\mathrm{L}},M_{\mathrm{H}}}(\vec{w}_{M_{\mathrm{H}}})}{\partial \vec{w}_{M_{\mathrm{H}}}} =
	\begin{pmatrix}
		0 & 0 & 0 & \cdots & 0 \\[5pt]
		a & 0 & 0 & \cdots & 0 \\[5pt]
		b_1 & 0 & c_{11} & \cdots & c_{1 M_{\mathrm{L}}} \\[5pt]
		\vdots & \vdots & \vdots & \ddots & \vdots \\[5pt]
		b_{M_{\mathrm{L}}} & 0 & c_{M_{\mathrm{L}} 1} & \cdots & c_{M_{\mathrm{L}} M_{\mathrm{L}}}
	\end{pmatrix},
\end{equation}
where we defined
\begin{equation}\label{eq:appendix_matrixDef}
	a = -\sum_{i=M_{\mathrm{L}}+1}^{M_{\mathrm{H}}}\frac{\alpha_i^2}{2i+1},\quad
	b_i = -2\sum_{j=1}^{M_{\mathrm{L}}}\sum_{k=M_{\mathrm{L}}+1}^{M_{\mathrm{H}}}A_{ijk}\alpha_j\alpha_k-\sum_{j,k=M_{\mathrm{L}}+1}^{M_{\mathrm{H}}}A_{ijk}\alpha_j\alpha_k,\quad
	c_{ij} = 2\sum_{k=M_{\mathrm{L}}+1}^{M_{\mathrm{H}}}A_{ijk}\frac{\alpha_k}{2k+1}.
\end{equation}

\paragraph{Non-conservative part.}
The mass balance equation \eqref{eq:SWME-h} and the momentum balance equation \eqref{eq:SWME-hum} do not contain non-conservative terms. The non-conservative term in the moment equation for \(\alpha_i\) \eqref{eq:SWME-halphai} is given by the term denoted by \(Q_{i+1}\) in \eqref{eq:SWME-halphai}. We write \(\delta \vec{Q}_{M_{\mathrm{L}},M_{\mathrm{H}}}(\vec{w}_{M_{\mathrm{H}}})\), which is defined via \(\delta\vec{Q}_{M_{\mathrm{L}},M_{\mathrm{H}}}(\vec{w}_{M_{\mathrm{H}}})=\delta Q_{M_{\mathrm{L}},M_{\mathrm{H}}}(\vec{w}_{M_{\mathrm{H}}})\partial_x \vec{w}_{M_{\mathrm{L}}}\), as
\begin{equation}\label{eq:diff_noncons_vectorentries}
	\delta \vec{Q}_{M_{\mathrm{L}},M_{\mathrm{H}}}(\vec{w}_{M_{\mathrm{H}}}) = (0,0,\delta Q_2,\ldots,\delta Q_{M_{\mathrm{L}}+1})^T.
\end{equation}
Subtracting the non-conservative flux terms \(Q_{i+1}\), \(i=1,\ldots,M_{\mathrm{L}}\), appearing in the moment equations for \(\alpha_i\) \eqref{eq:SWME-halphai}, \(i=1,\ldots,M_{\mathrm{L}}\), of the \(\text{SWME}_{M_{\mathrm{L}}}\) from the respective non-conservative flux terms in the first \(M_{\mathrm{L}}+2\) equations of the \(\text{SWME}_{M_{\mathrm{H}}}\), the entries \(\delta Q_{i+1}\) can be written explicitly for \(i=1,\ldots,M_{\mathrm{L}}\):
\begin{equation}\label{eq:diff_noncons}
	\delta Q_{i+1} = -\sum_{j=1}^{M_{\mathrm{L}}}\sum_{k=M_{\mathrm{L}}+1}^{M_{\mathrm{H}}}B_{ijk}\partial_x(h\alpha_j)\alpha_k  -\sum_{j=M_{\mathrm{L}}+1}^{M_{\mathrm{H}}}\sum_{k=1}^{M_{\mathrm{L}}}B_{ijk}\partial_x(h\alpha_j)\alpha_k-\sum_{j,k=M_{\mathrm{L}}+1}^{M_{\mathrm{H}}}B_{ijk}\partial_x(h\alpha_j)\alpha_k.
\end{equation}
Using \eqref{eq:diff_noncons}, the matrix \(\delta Q_{M_{\mathrm{L}},M_{\mathrm{H}}}(\vec{w}_{M_{\mathrm{H}}})\) can be constructed. We obtain
\begin{equation}\label{eq:diff_Q}
	\delta Q_{M_{\mathrm{L}},M_{\mathrm{H}}}(\vec{w}_{M_{\mathrm{H}}})=
	\begin{pmatrix}
		0 & 0 & 0 & \cdots & 0 \\[5pt]
		0 & 0 & 0 & \cdots & 0 \\[5pt]
		0 & 0 & -\sum_{k=M_{\mathrm{L}}+1}^{M_{\mathrm{H}}}B_{11k}\alpha_k & \cdots & -\sum_{k=M_{\mathrm{L}}+1}^{M_{\mathrm{H}}}B_{1M_{\mathrm{H}}k}\alpha_k \\[5pt]
		\vdots & \vdots & \vdots & \ddots & \vdots \\[5pt]
		0 & 0 & -\sum_{k=M_{\mathrm{L}}+1}^{M_{\mathrm{H}}}B_{M_{\mathrm{H}}1k}\alpha_k & \cdots & -\sum_{k=M_{\mathrm{L}}+1}^{M_{\mathrm{H}}}B_{M_{\mathrm{H}}M_{\mathrm{H}}k}\alpha_k
	\end{pmatrix}.
\end{equation}
The matrix \(A_{M_{\mathrm{H}}}(\vec{w}_{M_{\mathrm{H}}})\) is computed by inserting \eqref{eq:delta_F_full-matrix} and \eqref{eq:diff_Q} into \eqref{eq:systemMatr-x_Mh_appendix}.

Similarly, the quantity \( \delta \vec{S}_{M_{\mathrm{L}},M_{\mathrm{H}}}(\vec{w}_{M_{\mathrm{H}}})\) defined by \eqref{eq:A-S-decomposition2} can be calculated analytically. The vector \(\delta \vec{S}_{M_{\mathrm{L}},M_{\mathrm{H}}}(\vec{w}_{M_{\mathrm{H}}})\) is written as
\begin{equation}\label{eq:diff_source_vectorentries}
	\delta \vec{S}_{M_{\mathrm{L}},M_{\mathrm{H}}}(\vec{w}_{M_{\mathrm{H}}}) = (0,\delta S_0,\ldots,\delta S_{M_{\mathrm{L}}})^T \in \mathbb{R}^{M_{\mathrm{L}}+2}.
\end{equation}
Using the general SWME equations \eqref{eq:SWME-h}-\eqref{eq:SWME-halphai} and subtracting the source term of the \(\text{SWME}_{M_{\mathrm{L}}}\) from the first \(M_{\mathrm{L}}+2\) entries of the \(\text{SWME}_{M_{\mathrm{H}}}\), we obtain
\begin{equation}\label{eq:diff_source}
	\delta S_0 = -\frac{\nu}{\lambda}\sum_{j=M_{\mathrm{L}}+1}^{M_{\mathrm{H}}}\alpha_j, \quad \delta S_i = -(2i+1)\frac{\nu}{\lambda}\sum_{j=M_{\mathrm{L}}+1}^{M_{\mathrm{H}}}\left(1+\frac{\lambda}{h}C_{ij}\right)\alpha_j, \quad i=1,\ldots,M_{\mathrm{L}}.
\end{equation}
Inserting \eqref{eq:diff_source} in \eqref{eq:diff_source_vectorentries} yields \(\delta \vec{S}_{M_{\mathrm{L}},M_{\mathrm{H}}}(\vec{w}_{M_{\mathrm{H}}})\). It is readily observed from expressions \eqref{eq:delta_F_full-matrix}, \eqref{eq:diff_Q} and \eqref{eq:diff_source} that for vanishing moments \(\alpha_{M_{\mathrm{L}}+1}=\ldots=\alpha_{M_{\mathrm{H}}}=0\), both \(A_{M_{\mathrm{H}}}(\vec{w}_{M_{\mathrm{H}}})\) and \(\delta \vec{S}_{M_{\mathrm{L}},M_{\mathrm{H}}}(\vec{w}_{M_{\mathrm{H}}})\) vanish.


\end{document}